\documentclass[%
reprint,
superscriptaddress,
 amsmath,amssymb,
 aps,
prb,
]{revtex4-2}

\usepackage{graphicx}% Include figure files
\usepackage{dcolumn}% Align table columns on decimal point
\usepackage{bm}% bold math
\usepackage{color}
\usepackage{cancel}
\usepackage{comment}
\usepackage{appendix}
\usepackage{braket}
\allowdisplaybreaks
\usepackage[colorlinks=true,
  allcolors={blue}]{hyperref}% add hypertext capabilities
\usepackage{cleveref}
\begin{document}

\title{Anisotropic effects in two-dimensional materials}

\author{Alexander N. Rudenko}
\email{a.rudenko@science.ru.nl}

\author{Mikhail I. Katsnelson}
\affiliation{\mbox{Radboud University, Institute for Molecules and Materials, Heijendaalseweg 135, 6525AJ Nijmegen, The Netherlands}}

\date{\today}

\begin{abstract}
Among a huge variety of known two-dimensional materials, some of them have anisotropic crystal structures; examples include so different systems as a few-layer black phosphorus (phosphorene), beryllium nitride BeN$_4$, van der Waals magnet CrSBr, rhenium dichalcogenides ReX$_2$. As a consequence, their optical and electronic properties turn out to be highly anisotropic as well. In some cases, the anisotropy results not just in a smooth renormalization of observable properties in comparison with the isotropic case but in the appearance of dramatically new physics. The examples are hyperbolic plasmons and excitons, strongly anisotropic ordering of adatoms at the surface of two-dimensional or van der Waals materials, essential change of transport and superconducting properties. Here, we present a systematic review of electronic structure, transport and optical properties of several representative groups of anisotropic two-dimensional materials including semiconductors, anisotropic Dirac and semi-Dirac materials, as well as superconductors. 
\end{abstract}

\maketitle

%\tableofcontents

\section{Introduction}

The discovery of graphene in 2004 \cite{Novoselov2004} has paved the way to a whole new world of two-dimensional (2D) materials with highly diverse physical properties ranging from semiconductors and normal metals to more exotic forms of matter such as superconductors, magnets, and topologically nontrivial systems \cite{Novoselov2016,Ren2016,Saito2016,Gibertini2019}. On the one hand, the existence of truly 2D materials is important to understand and probe the fundamental laws of physics at low dimensions, which give rise to the emergence of qualitatively new effects not present in the 3D world. On the other hand, 2D materials offer great tunability of their properties  by means of external stimuli, which opens up broad prospects for applications, especially in the field of electronics and optics \cite{Fiori2014,Shaoliang2017,Liu2019_review,Glavin2020}. The ability to combine various 2D material to form heterostructures with controllable stacking and misorientation \cite{Geim2013,Andrei2020} is a truly inexhaustible and one of the most promising directions of today's material science.

Most of the known 2D materials adopt crystal lattices with hexagonal symmetry, making many of their properties isotropic at the macro scale. 
Strictly speaking, none of the crystalline materials are perfectly isotropic because of the absence of full rotational symmetry. It is well known that translation invariance of conventional crystals is only compatible with 2-, 3-, 4-, and 6-fold rotations, ensuring that the electronic structure is in general anisotropic at finite wave vector \cite{vk_book}.
However, macroscopic properties described by rank-2 tensors, such as dielectric permittivity $\varepsilon_{\alpha \beta}(\omega)$, are isotropic for highly symmetric (including hexagonal) crystals, i.e., $\varepsilon_{\alpha \beta}(\omega) = \varepsilon(\omega) \delta_{\alpha \beta}$ \cite{Landau_ED}. On the other hand, low symmetry crystals lacking 3-, 4-, and 6-fold rotations are anisotropic even at the macro scale, for which $\varepsilon_{xx}(\omega) \neq \varepsilon_{yy}(\omega)$. These specific crystals constitute the subject of the present Review.

Some 2D materials indeed exhibit highly anisotropic crystal structures, giving rise to anisotropy of their observable properties. The most studied 2D material of this kind is black phosphorus (BP), a single layer of which was dubbed phosphorene, in analogy to graphene. The 3D BP is the most stable allotrope of phosphorus, which crystallizes in the so-called A17 structure, an orthorhombic structure consisting of weakly bounded layers. Few-layer BP was first mechanically exfoliated in 2014 \cite{Li2014,Liu2014,Xia2014,Koenig2014,Buscema2014}, which attracted enormous attention from the 2D community \cite{Dresselhaus2015}, and essentially triggered theoretical studies of 2D anisotropic materials. It is natural to ask: Does anisotropy play any special role in the formation of physical properties of 2D materials, or does it simply govern their direction dependence?

In the present Review, we address the question formulated above in the context of electronic and optical properties. We restrict ourselves mostly to theoretical aspects of the 2D anisotropy starting from basic electronic structure models up to the state-of-the-art first-principles calculations. We attempt to demonstrate that the anisotropy effects are not limited to the appearance of quantitative corrections to the angular dependence of observables, but they may be the origin of qualitatively new phenomena, not having isotropic analogs. The most prominent example is the emergence of a special type of electromagnetic waves propagating in 2D anisotropic materials known as hyperbolic polaritons, which were originally analyzed in the context of electromagnetic metamaterials \cite{Poddubny2013,Shekhar2014}. Another example is related to electrostatic interactions at the atomic scale, giving rise to the anisotropic arrangement of adatoms on anisotropic surfaces, opening up perspectives for creating neuromorphic architectures \cite{Kiraly2021}. Essential anisotropy of the energy spectrum also manifests itself in a significant modification 
of the materials' response to external fields, resulting in qualitatively different power-law dependencies. From this perspective, anisotropic 2D systems can be viewed as an intermediate form of matter between isotropic 2D and quasi-1D. 
Last but not the least, anisotropic character of materials' properties requires a careful reformulation of the conventional \cite{mermin_book,kittel_book,Ziman-Book,vk_book} theory approaches. Some well-known expressions commonly used to analyze properties of isotropic materials may become highly inaccurate even for moderately anisotropic systems. From the application viewpoint, the anisotropy itself can be exploited as an additional degree of freedom, whose controllable manipulation may result in the emergence of desired qualities.

It is hardly possible to cover all aspects of anisotropic 2D materials in one review, so the present Review does not pretend to be complete. For example, we intentionally excluded magnetic materials from the consideration as we believe anisotropic aspects of magnetism represent a stand-alone topic, deserving a separate review. We also do not consider mechanical properties of anisotropic membranes, which is an interesting but not extensively studied topic \cite{Burmistrov2022,Parfenov2022}.
Thermodynamical properties of anisotropic 2D materials is another topic that is perhaps too early to summarize. By analogy with 3D materials \cite{licht_jones1,licht_jones2,titanium} one can expect, for example, bright anomalies in anisotropic thermal expansion but we are not familiar with any experimental or theoretical considerations of this issues for 2D case, so let us remain this subject for the future.

The main part of our Review is organized in three sections as follows. In Section \ref{sec2}, we provide a classification of 2D materials with respect to their electronic structure, and overview basic models. Specifically, we distinguish between (i) anisotropic semiconductors with quadratic dispersion relation, (ii) Dirac semimetals with linear dispersion yet anisotropic Dirac cone, and (iii) semi-Dirac semimetals that exhibit a hybrid linear/quadratic dispersion.
Section \ref{sec3} is devoted to the analysis of dielectric screening and optical response of 2D anisotropic materials. In Sec.~\ref{sec3a}, we analyze optical conductivity, and highlight unusual behavior originating from their anisotropic character. The peculiarities of 2D anisotropic screening, dynamical polarizability and plasmons are discussed in Sec.~\ref{plasmons}, where we pay a special attention to the hyperbolic regime of plasmon propagation. Anisotropic effects related to the excitons in 2D materials are reviewed in Sec.~\ref{sec3c}. In Section \ref{sec4}, we discuss charge carrier scattering and electronic transport. We first provide basic semiclassical theory of anisotropic transport (Sec.~\ref{theory_scatter}), and then consider specific scattering mechanisms, including scattering by random disorder (Sec.~\ref{sec4b}), charge impurity scattering (Sec.~\ref{sec_chg-imp}), and electron-phonon scattering (Sec.~\ref{sec4d}). 
In Sec.~\ref{sec4e}, we discuss the role of anisotropy in superconductivity, and demonstrate the role of proximity-induced anisotropy in conventional superconductors. Our Review is concluded by a summary and outlook.

\section{\label{sec2}Electronic properties and basic models}

\subsection{\label{sec2a}Anisotropic 2D electron gas}
\begin{figure}[t]
\centering
\mbox{
\includegraphics[width=1.00\linewidth]{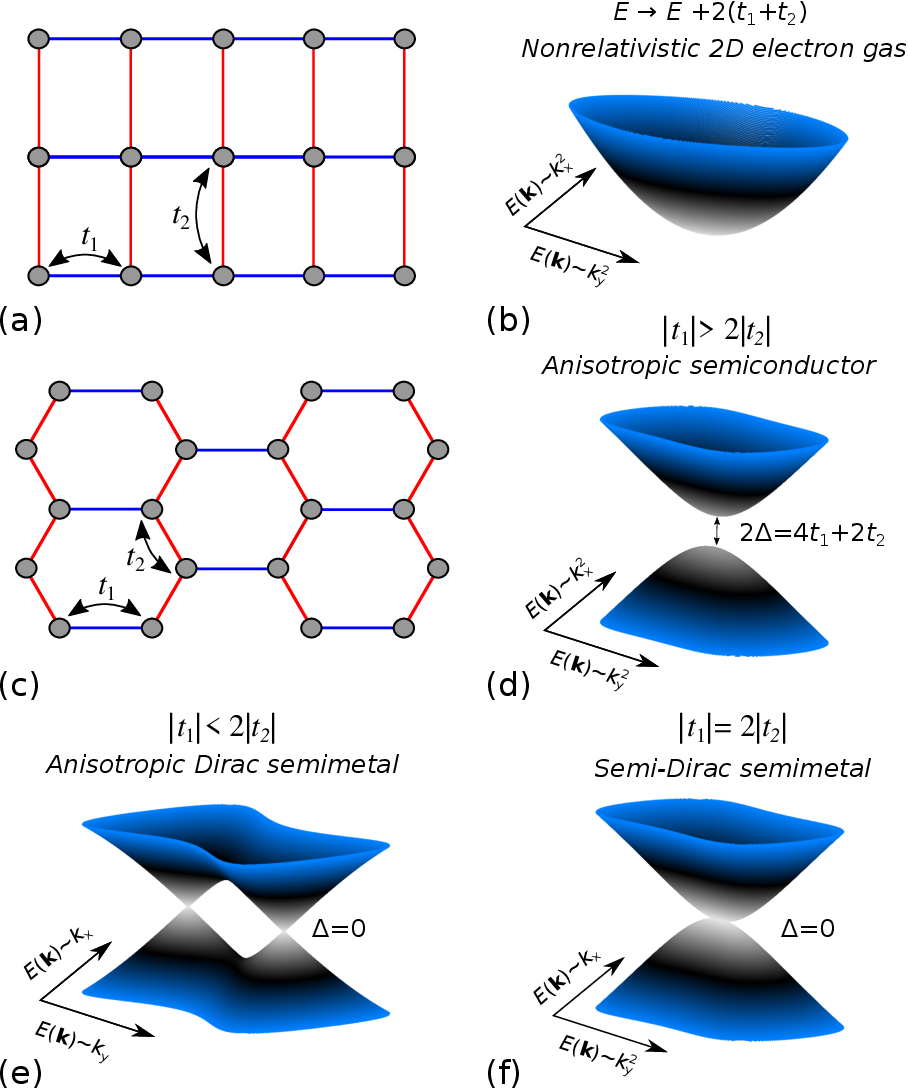}
}
\caption{(a) A sketch of the rectangular lattice with two nearest-neighbor hoppings; (b) Band structure of anisotropic 2D electron gas near the lowest band edge; (c) Distorted honeycomb lattice with two nearest-neighbor hoppings; (d) Band structure of a 2D anisotropic semiconductor around the band gap; (e) Band structure of anisotropic 2D Dirac semimetal; (f) Band structure of a semi-Dirac semimetal.}
\label{fig1}
\end{figure}

One of the simplest models of an anisotropic material is given by one-particle tight-binding (TB) Hamiltonian of the form,
\begin{equation}
    H = \sum_{ij}t_{ij}a^{\dag}_i a_j,
    \end{equation}
where $a^{\dag}_i$ and $a_j$ are the creation and annihilation operators of electrons at sites $i$ and $j$, and $t_{ij}$ is the corresponding hopping integral, quantifying the energy to move an electron from site $j$ to site $i$. Schematically, the TB model on a rectangular lattice is shown in Fig.~\ref{fig1}(a). In reciprocal space, single-particle representation of the above Hamiltonian takes the form,
\begin{equation}
H({\bf q}) = 2t_1 \, \mathrm{cos}(q_xa) + 2t_2 \, \mathrm{cos}(q_yb),   
\label{tb-rect}
\end{equation}
which describes a single band of the width $W=4(t_1+t_2)$.
%, with the band edges located at the $\Gamma$ point. 
Expanding $H({\bf q})$ around ${\bf q}=(\pi,\pi)$ and keeping the quadratic terms in ${\bf q}$ only, we arrive at the energy dispersion of anisotropic 2D electron gas (2DEG)
\begin{equation}
  E({\bf q}) - E_0 = \frac{\hbar^2q_x^2}{2m_x} + \frac{\hbar^2q_y^2}{2m_y},
    \label{el-gas}
\end{equation}
where $m_x = -\hbar^2/2t_1a^2$ and $m_y = -\hbar^2/2t_2b^2$ are the effective masses, and $E_0 = -W/2$ is the energy of the lowest band edge. The corresponding density of states (DOS) per spin is given by $g(E) = \sqrt{m_x m_y}/(2\pi \hbar^2$). Within this approximation, the Fermi contour is represented by an ellipse. The corresponding dispersion relation $E({\bf k})$ is shown schematically in Fig.~\ref{fig1}(b).

Realization of 2DEG does not necessarily involve 2D materials. Instead, charge carrier can be confined at the interface between insulating materials, giving rise to a highly conducting 2DEG, with the typical representative being  LaAlO$_3$/SrTiO$_3$ \cite{Ohtomo2004}. Depending on the crystal orientation and interface morphology, anisotropic 2DEG can be realized. For example, Brinks \emph{et al.} \cite{Brinks2011} reported strong anisotropy of the electron gas confined at the SrTiO$_3$–LaAlO$_3$ interface, which was observed by measuring angular dependence of the carrier mobility. The anisotropy origin was ascribed to the presence of step edges at the interface. A similar strong anisotropy of the electronic transport at the same heterointerface was observed by Annadi \emph{et al.} \cite{Annadi2013}, which was attributed to a buckling of the interface atomic structure. Later, by performing angle-resolved photoemission spectroscopy (ARPES) experiments, Wang \emph{et al.} \cite{Wang2014} observed anisotropic electronic structure of 2DEG at the (110) surface of SrTiO$_3$, which undergoes reconstruction, forming a structurally anisotropic Ti overlayer. Other heterostructures with observed anisotropic electronic structure include GaN/GaAlN \cite{Lev2018}, AlN/GaN \cite{Yang2019}, AlO$_x$/KTaO$_3$(110) \cite{Martinez2023}.

\subsection{\label{sec2b}Anisotropic semiconductors}
In the vicinity of the band edges, the model Eq.~(\ref{tb-rect}) describes anisotropic nonrelativisic 2D electron gas, which can be realized in an anisotropic semiconductor by either electron or hole doping, i.e. upon filling of either valence or conduction band. In order to describe the valence and conduction bands on equal footing, the single-band model given by Eq.~(\ref{tb-rect}) is obviously insufficient. A more general TB description of anisotropic materials can be constructed on the basis of a distorted honeycomb lattice [Fig.~\ref{fig1}(c)]. This model has been analyzed by Pereira \emph{et al.} \cite{Pereira2009} in the context of strained graphene. However, as demonstrated by Montambaux \emph{et al.} \cite{Montambaux2009,Montambaux2009-2} it has more general applicability. At the same time, we note that the honeycomb lattice is by far not the only choice to approach the
electronic structure of anisotropic 2D materials, including Dirac semimetals. Other lattices such as, for instance, rectangular or triangular lattices with a basis provide an equally possible choice \cite{Miert2016}. 
%Nevertheless, we believe that the honeycomb lattice is more convenient for pedagogical reasons.  

The TB Hamiltonian for a honeycomb lattice can be written as
\begin{equation}
H({\bf q})=
    \begin{pmatrix}
        0 & f({\bf q}) \\
        f^*({\bf q}) & 0 
        \end{pmatrix},
    \label{tb-honey}
\end{equation}
where
\begin{equation}
    f({\bf q}) = 2t_1\,\mathrm{exp}\left(\frac{iq_xb}{2}\right)\mathrm{cos}\left(\frac{\sqrt{3}}{2}q_yb\right) + t_2\,\mathrm{exp}\left(-iq_xa\right).
    \label{f_q}
\end{equation}
The energy dispersion is given by $E({\bf q}) = \pm |f({\bf q})|$. If $E({\bf q})=0$, the two bands have a crossing at
\begin{equation}
    {\bf q}_0 = \left[ \frac{2\pi}{2a+b} ; \, \pm\frac{2}{\sqrt{3}b}\mathrm{ arccos}\left( \frac{t_2}{2t_1}\right) \right]. 
    \label{q0}
\end{equation}
If $t_1=t_2$ and $a=b$ (undistorted honeycomb lattice), the ${\bf q}_0$ points correspond to the K and K' points of the hexagonal Brillouin zone, as in graphene \cite{Katsnelson-Book}. In the situation when $|t_2|>2|t_1|$, there is no crossing between the valence and conduction bands, and the model Eq.~(\ref{tb-honey}) describes anisotropic semiconductor with an energy gap 2$\Delta = 4t_1+2t_2$ at the $\Gamma$ point, showing schematically in Fig.~\ref{fig1}(d).

In the low-energy limit at $|t_2|>2|t_1|$ (gapped regime), the Hamiltonian Eq.~(\ref{tb-honey}) can be expanded up to the second order in ${\bf q}$ as
%(in what follows, we assume $\hbar=1$ for brevity)
\begin{equation}
H =
    \begin{pmatrix}
        0 & \Delta + \frac{\hbar^2 q_x^2}{2m_x} + \frac{\hbar^2 q_y^2}{2m_y} - i\hbar v_xq_x \\
        \Delta + \frac{\hbar^2 q_x^2}{2m_x} + \frac{\hbar^2 q_y^2}{2m_y} + i\hbar v_xq_x & 0 
        \end{pmatrix},
    \label{H_expand}
\end{equation}
where $m_x^{-1} = -3t_1b^2/2\hbar^2$, $m_y^{-1} = -(t_1b^2/2 + t_2a^2)/\hbar^2$, and $v_x = (t_1b - t_2a)/\hbar$. Surprisingly, the Hamiltonian acquires a linear term in the $x$ direction, being a consequence of the symmetry lowering, and which vanishes if $t_1=t_2$ and $a=b$. The low-energy energy spectrum then reads
\begin{equation}
    E ({\bf q}) = \pm \sqrt{\left(\Delta + \frac{\hbar^2 q_x^2}{2m_x} + \frac{\hbar^2 q_y^2}{2m_y}\right)^2 + \hbar^2 v_x^2q_x^2},
    \label{E-q}
\end{equation}
shown schematically in Fig.~\ref{fig1}(d).
One can immediately see that the electronic properties of anisotropic lattices are generally different from those of anisotropic 2DEG.
Nevertheless, one can estimate the effective masses from Eq.~(\ref{E-q}), $(m^*_i)^{-1} = \hbar^{-2}\partial^2E({\bf q})/\partial q^2_i$ ($i=x,y$), keeping in mind limited applicability of the effective mass approximation away from the band edges. Along the $y$ direction, we have $m^*_y=m_y$, whereas along $x$ the effective mass is corrected by the linear term, resulting in $(m_x^*)^{-1} = m_x^{-1} + v_x^2/\Delta$. In the case $v_x^2/\Delta \gg m_x^{-1}$, e.g. when $\Delta$ is small, DOS per spin can be expressed for $E>\Delta$ as \cite{Montambaux2009-2,Zyuzin2015}
\begin{equation}
g(E) = \frac{\sqrt{m_y}}{2\pi^2\hbar^2 v_x}\sqrt{E}\,K\left(\sqrt{\frac{E-\Delta}{2E}}\right),
\label{dos_ellipt}
\end{equation}
where $K(x)$ is the complete elliptic integral of the first kind. One can see that in contrast to the energy-independent DOS of 2DEG, the presence of the linear term $\sim$$ v_xq_x$ in the Hamiltonian Eq.~(\ref{H_expand}) results in a square-root dependence of DOS. The resulting spectrum is particle-hole symmetric $g(E)=g(-E)$, which is governed by the sublattice symmetry of the lattice model [Fig.~\ref{fig1}(c)]. In case of real materials, the particle-hole symmetry can be broken by adding more hoppings within the same sublattice in the model.

In the presence of an external magnetic field ${\bf B}$ the energy spectrum can be calculated by considering the Peierls substitution ${\bf p} \rightarrow {\bf p} - e{\bf A}$ \cite{vk_book,Katsnelson-Book} in the Hamiltonian Eq.~(\ref{H_expand}). In case of a magnetic field perpendicular to the 2D plane ${\bf B}=(0,0,B)$, and using the Landau gauge ${\bf A} = (-By,0,0)$, the spectrum of Landau levels can be obtained. For sufficiently large $\Delta/B^{2/3}$, the resulting Hamiltonian can be expanded and the usual Landau spectrum is recovered \cite{Montambaux2009-2},
\begin{equation}
    E_n = \pm \left(  \Delta + (n + 1/2) \hbar \omega_\mathrm{c} \right),
    \label{landau}
\end{equation}
where $\omega_\mathrm{c}=eB/\sqrt{m^*_xm^*_y}$ is the cyclotron frequency, and each level is nondegenerate. In the situation, when the condition $\Delta/B^{2/3} \gg 1$ is not fulfilled, (e.g., at high enough fields), a sublinear dependence on the level index $n$ is realized \cite{Yuan2015,Milton2015}. For realistic anisotropic semiconductors (e.g., black phosphorus), this regime is barely achievable, making Eq.~(\ref{landau}) applicable in most of the practical situations.

For a multilayer system with a nonzero interaction between the layers, the valence and conduction band split into subbands, with the splitting proportional to the interlayer hopping $t_{\perp}$ \cite{Zyuzin2015,Rudenko2015,Souza2017}. In case of a bilayer, the energy gap becomes $2\Delta^{(2)} = 2(2t_1+t_2\sqrt{1+(t_{\perp}/t_2)^2} - t_{\perp})$. Under certain assumptions (e.g., $t_{\perp}/t_2 \ll 1$), the gap for a $N$-layer system can be written as \cite{Souza2017}
\begin{equation}
2\Delta^{(N)} = 2\left(2t_1+t_2 - 2\,\mathrm{cos}\left(\frac{\pi}{N+1}\right)t_{\perp}\right).    
\end{equation}

\begin{figure}[t]
\centering
\mbox{
\includegraphics[width=0.80\linewidth]{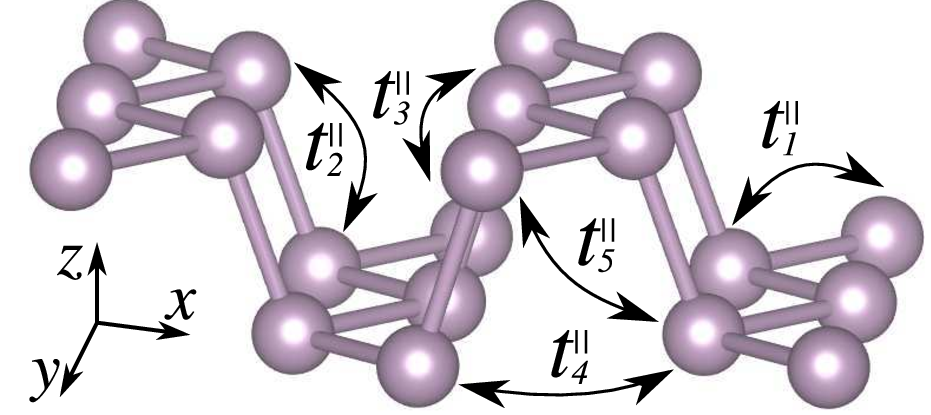}
}
\caption{Schematic atomic structure of monolayer BP with the main hoppings from the TB model proposed in Refs.~\onlinecite{Rudenko2014,Rudenko2015}.}
\label{fig2}
\end{figure}

The most well-known example of anisotropic 2D semiconductors is the 2D derivative of orthorhombic A17 phase of black phosphorus (BP). The experimental exfoliation of a few-layer BP dates back to 2014 \cite{Li2014,Xia2014,Koenig2014,Buscema2014}. BP is a layered semiconductor in which each layer adopts a puckered structure with finite thickness.
The lattice structure of black phosphorus is different from the distorted honeycomb lattice [Fig.~\ref{fig2}], giving rise to a four-band TB model for monolayer BP, which is slightly more complicated compared to Eq.~(\ref{tb-rect}) \cite{Rudenko2014,Rudenko2015}. Nevertheless, as has been shown by Ezawa \cite{Ezawa2014}, the four-band TB model can be reduced to the two-band TB model of a distorted honeycomb lattice
due to the $D_{\mathrm{2h}}$ point group invariance. The two-band TB model assumes zero thickness of each layer and satisfactorily describes the low-energy properties of monolayer BP in the situations when real-space description is desirable (e.g., for nanoribbons), while layer thickness is not relevant.

The effective masses in BP are considerably different between the two in-plane crystallographic directions, commonly referred to as armchair ($x$) and zigzag ($y$). For both electrons and holes in monolayer as well as multilayer BP, $m_x/m_y \ll 1$, being a consequence of the linear term along $x$ in the Hamiltonian. First-principles calculations within the $GW_0$ method predict $m^{N=1}_x\approx 0.2$ (in the units of free electron mass $m_0$) for monolayer BP, which gradually decrease to $m^{\infty}_x\approx 0.1$ in bulk BP for both electrons and holes \cite{Rudenko2015,Souza2017}. Along the zigzag direction, the effective masses are more sensitive to the number of layers, displaying an opposite trend for electrons and holes. Specifically, the hole mass decreases from  $m^{N=1}_y\approx 3.1m_0$ in the monolayer to $m^{\infty}_y\approx 0.8m_0$ in the bulk, while the electron mass increases from $1.0m_0$ to $1.9m_0$, respectively. It should be noted that the state-of-the-art first-principles calculations within the quasiparticle self-consistent $GW$ approach with vertex corrections in the electronic channel predict somewhat smaller effective masses in bulk BP \cite{Rudenko2022}, yet showing similar trends.

Contrary to phosphorus, other group-V elements tend to crystallize with the rhombohedral A7 structure, which is the most stable allotropic form of As, Sb, and Bi at ambient conditions \cite{Pumera2017,Sofer2020}. While a stable form of orthorhombic As (also referred to as $\alpha$-phase) exists and can even be mechanically exfoliated to a few-layer crystals \cite{Zhong2018,Chen2018}, the heavy pnictogens Sb and Bi do not form the A17 phase in nature. Being unstable in bulk, such phases can still be synthesized in the form of a few-layer crystals using epitaxial growth on surfaces \cite{Mannix2017,Shi2019}. Also, it is interesting to note a recent experimental discovery of crystalline nitrogen with the A17 structure at extremely high pressure of $\sim$140 GPa \cite{Ji2020,Laniel2020}. Although the electronic properties of this N allotrope are in many respects similar to that of black phosphorus \cite{Rudenko2022}, its 2D counterpart would be unstable even in theory.

Among the large number of layered transition metal dichalcogenides, ReS$_2$ and ReSe$_2$ are known as typical examples of materials with in-plane anisotropy, resulting from Peierls distortion of the 1$T$ structure with the triclinic symmetry. ARPES measurements reveal a considerable difference in the effective masses between in-plane crystallographic directions, with the ratio reaching a factor of two \cite{Tongay2014,Hart2017}, which is in line with first-principles calculations \cite{YungChang2015}. Another group of 2D anisotropic semiconducting materials is represented by the monoclinic phases of group IV transitional metal trichalcogenides, with TiS$_3$ and ZrS$_3$ being experimentally available \cite{Island2015,Pant2016}. Apart from being strongly anisotropic, these materials are characterized by opposite ellipticity of the valence and conduction bands \cite{Jose2017}, which is uncommon to other materials. To the family of experimentally available 2D semiconductors with highly anisotropic electronic and optical properties one can also append orthorhombic monochalcogenides of group V elements (e.g., GeSe) \cite{Tan2017,Wang2017}, their monoclinic monopnictides (e.g., GeAs) \cite{Guo2018}, as well as orthorhombic As$_2$S$_3$ \cite{Siskin2019,Liu2021,slavich2023exploring}.

\subsection{\label{sec2c}Anisotropic Dirac semimetals}

Let us now return to the distorted honeycomb lattice [Fig.~\ref{fig1}(c)] with the Hamiltonian Eq.~(\ref{tb-honey}), and consider the situation when $|t_2|<2|t_1|$. In this case, we have two crossing points ${\bf q}_0$ in the Brillouin zone given by Eq.~(\ref{q0}), see Fig.~\ref{fig1}(e). In the vicinity of the crossing points, the Hamiltonian can be linearized, yielding
\begin{equation}
    H =
        \begin{pmatrix}
        0 & \pm \hbar v_y(q_y-q_{0y}) - i\hbar v_xq_x \\
        \pm \hbar v_y(q_y-q_{0y}) + i\hbar v_xq_x & 0 
        \end{pmatrix},
        \label{H-anisD}
\end{equation}
where $v_x$ and $v_y=\hbar q_{y0}/m_y$ play the role of the Fermi velocities along the $x$ and $y$ directions, respectively.
The corresponding energy dispersion reads
\begin{equation}
    E({\bf q}) = \pm\sqrt{\hbar^2 v^2_xq_x^2 + \hbar^2 v_y^2(q_y - q_{y0})^2},
\end{equation}
with the constant-energy contours having an elliptical shape. In the low-energy limit, DOS per spin is given by $g(E) = |E| / (2\pi\hbar^2 v_x v_y)$, which is reminiscent of DOS for graphene within the Dirac cone approximation. A more general expression, beyond this limit, can be found, e.g., in Ref.~\onlinecite{Montambaux2009-2}. It is interesting to note that for an $N$-layer anisotropic Dirac material one has $g(E)\propto 1/(N|E|)$ at $N\gg1$ \cite{Zyuzin2015}, which is different from the behavior of multilayer graphene \cite{Katsnelson-Book}, where $g(E) \propto 1/|E|^{1-2/N}$.

In the presence of a perpendicular magnetic field $B$, the Landau spectrum of anisotropic Dirac semimetal depends on the parameter $\delta \propto  v_yq_{y0}/B^{2/3}$. For $\delta\gg 1$, one obtains the form,
\begin{equation}
    E_n = \pm\sqrt{2ne\hbar v_xv_yB},
\end{equation}
typical for the Dirac materials. Here, each level except $n=0$ is double degenerate, as a result of the two independent Dirac valleys \cite{Montambaux2009,Montambaux2009-2}. If the condition $\delta\gg 1$ is not satisfied, the potential barrier between the valley decreases, lifting the valley degeneracy and modifying the $E_n\propto \sqrt{nB}$ dependence toward a more linear one.

The most natural way to realize an anisotropic
Dirac semimetal is to apply uniaxial strain to graphene. This problem was studied by Pereira \emph{et al.} \cite{Pereira2009} at the level of a TB model with renormalized hopping parameters. It was demonstrated that a strain of 5\% visibly deforms the Fermi contour of graphene, resulting in anisotropic Fermi velocities, with the ratio $v_x/v_y\sim1.2$ for a Fermi energy of 50 meV. Experimentally, strain in graphene can naturally occur as a result of the lattice mismatch with the substrate. In most of the situations, however, this leads to biaxial strain. Uniaxial strain might occur in graphene grown on substrates with anisotropic morphology (e.g., step edges). In this case, the Fermi velocity modification mostly occur along the direction of the lattice deformation, which can be directly measured by means of ARPES \cite{Kan2012}. Such methods do not provide wide opportunities for anisotropy control.

Another approach to induce anisotropy in the electronic structure of a Dirac material is to apply periodically modulated potentials to graphene, creating a superlattice \cite{Park2008,Tianlin2024}. These potentials may originate from periodic deformation fields, gate-voltage modulation, controllable adatom deposition, \emph{etc}. With these approaches, much higher anisotropies can be achieved, only limited by the potential strength and its periodicity \cite{Park2008}. Interestingly, for a 1D potential modulation, the Fermi velocity is reduced mostly in the direction perpendicular to the periodicity of the potential. Moreover, the potential parameters could be tuned in such a way that the Fermi velocity vanishes in one of the directions, making the system effectively 1D.

The theory of graphene in generic periodic deformation fields has been developed in Ref.~\onlinecite{Dugaev2012}. It
is known that the deformation of a honeycomb lattice is equivalent, for a given valley, to the generation of electric and magnetic fields \cite{Katsnelson-Book}, which can be
described by scalar 
\begin{equation}
	V({\bf r})=g[u_{xx}({\bf r}) + u_{yy}({\bf r})],    
\end{equation}
and vector 
\begin{eqnarray}
\notag
A_x({\bf r}) = \frac{\beta t}{a}[u_{xx}({\bf r}) - u_{yy}({\bf r})], \\
A_{y}({\bf r}) = -\frac{2\beta t}{a}u_{xy}({\bf r}) \quad \quad \quad \,
\end{eqnarray} 
potentials. Here, $u_{ij}({\bf r})$ is the strain tensor, $g$ is the deformation potential, $t$ is the nearest-neighbor hopping of an unperturbed lattice, $a$ is the lattice constant, and $\beta=-\partial \, \mathrm{ln}\,t / \partial \, \mathrm{ln} \,a$. The presence of a 1D periodic vector field in the form ${A}_y(x)=A_0\,\mathrm{sin}(2\pi \,x/L)$ does not induce anisotropy of the Fermi velocity, but only leads to its isotropic reduction, which is equivalent to the results obtained for graphene under a 1D inhomogeneous magnetic field \cite{Tan2010}. On the contrary, under the scalar periodic fields, the components of the Fermi velocity are renormalized unequally, leading to the emergence of the electronic anisotropy. For the scalar potential $V(x)$ periodic in the $x$ direction, the components of the group velocity can be written as $v_x = v\sqrt{1 + \gamma_1^2}$ and $v_y = v |\gamma_2|$, where $\gamma_1$ and $\gamma_2$ are real and imaginary components of the quantity
\begin{equation}
    \gamma = \frac{1}{L}\int_0^L dx \, \, \mathrm{exp}\left[ \frac{2i}{v}\int_0^x dx' \, V(x') \right],
\end{equation}
with $L$ being the period of the potential $V(x)$. One can see that $v_x>v$ and $v_y<v$.

The consideration given above is based on single-particle picture without many-body corrections. It is worth noting that the dispersion of massless 2D Dirac electrons in the presence of long-range Coulomb interactions undergoes a strong renormalization which is formally logarithmically divergent at zero temperature and doping \cite{Katsnelson-Book}. 
In the context of anisotropic graphene, this problem was first addressed by Dugaev and Katsnelson in Ref.~\onlinecite{Dugaev2012}.
They obtained the following renormalization group flow equations for the renormalized velocities based on the first-order Hartree-Fock approximation:
\begin{eqnarray}    
\label{Dugaev_eq}
\notag
\frac{\partial \tilde{v}_x}{\partial \xi} = \frac{e^2 }{\pi \hbar } \frac{\tilde{v}_x}{\tilde{v}_y}\left[ \mathrm{K}(m) - \frac{\tilde{v}_x^2}{\tilde{v}_y^2}\mathrm{R}(m) \right],\\
\frac{\partial \tilde{v}_y}{\partial \xi} =   \frac{e^2}{\pi \hbar} \left[ \mathrm{K}(m) - \mathrm{P}(m)\right], \quad \, \, \,
\end{eqnarray}
where $\xi = \mathrm{log}(\Lambda/k_\mathrm{F})$, $\Lambda$ is the running cut-off momentum parameter, and $\textsf{K}(m)=\int _0^{\pi /2}(1-m\sin ^2\theta )^{-1/2}d\theta$,
$\textsf{P}(m)=\int _0^{\pi /2}(1-m\sin ^2\theta )^{-3/2}
\sin ^2\theta \, d\theta $,
$\textsf{R}(m)=\int _0^{\pi /2}(1-m\sin ^2\theta )^{-3/2}
\cos ^2\theta \, d\theta $, and $m=1-\tilde{v}_x^2/\tilde{v}_y^2$. The solution of Eq.~(\ref{Dugaev_eq}) indicates that the renormalized energy spectrum is always less anisotropic than the bare one. In fact, Leaw \emph{et al.} \cite{Leaw2019} have noticed that the solutions of Eq.~(\ref{Dugaev_eq}) can be quite reasonably approximated by a simple square-root renormalization of the velocity anisotropy, i.e. $\tilde{v}_x/\tilde{v}_y \approx \sqrt{v_x/v_y}$.  
Based on a more accurate many-body consideration including lattice quantum Monte Carlo computations, it was shown that this result is universal and unique to Dirac fermions in 2D \cite{Leaw2019}, independent of the interaction strength. Therefore, the interacting states in graphene-like lattices are always more isotropic compared to the noninteracting single-particle states. This finding is reminiscent to the behavior of the quantum Hall states at half filling, such as the composite fermions, where a similar square-root renormalization was observed \cite{Jo2017,Ippoliti2017}. 

Another possible approach in the realization of anisotropic Dirac dispersion relies on the tunability of a band gap in narrow-gap anisotropic semiconductors, artificially inducing a semiconductor-semimetal transition by means of mechanical strain, external fields, or doping. For this purpose, anisotropic BP, layered semiconductor with a $\sim$0.3 eV energy gap in the bulk phase, appears like an excellent candidate. Indeed, using first-principles calculations Fei \emph{et al.} \cite{Fei2015} showed that bulk or multilayer BP undergoes a transition to the gapless state at the pressure of 0.6 GPa, forming topologically protected 2D Dirac cones at higher pressures. Similar results were obtained by Dutreix \emph{et al.} \cite{Dutreix2016} considering in-plane time-periodic laser fields with linear polarization in the context of single-layer BP. Experimentally, tunability of the band gap in BP was demonstrated by Kim \emph{et al.} \cite{Kim2015}, who doped few-layer BP by K atoms and found a transition to the anisotropic Dirac semimetal state, dependent on the
K concentration. The effect was ascribed to the giant Stark effect induced by the ionized K atoms deposited on the surface. First-principles calculations with explicit deposition of K atoms \cite{Baik2015}, as well as in the presence of vertical static electric field \cite{Zunger2015} have confirmed this scenario.

A conceptually different approach for the realization of Dirac fermions is offered by artificial honeycomb lattices \cite{Polini2013}. Such structures may be created by trapping of ultracold atoms in optical lattices \cite{Zhu2007}, molecular structures assembled on surfaces by atomic manipulation \cite{Gomes2012}, and confined photons in photonic crystals \cite{Plotnik2014}.
The tunability and manipulation of the Dirac points, including the formation of anisotropic Dirac cones and semi-Dirac points, were proved to be possible in an ultracold Fermi gas of $^{40}$K atoms in a 2D optical honeycomb lattice 
\cite{Tarruell2012}.

The number of intrinsically anisotropic and truly 2D Dirac semimetals is not large and limited to a few examples. Apart from borophenes, which will be discussed later in this section, the theoretically predicted system with slightly anisotropic Dirac cones include honeycomb monolayers TiB$_2$ \cite{Zhang2014} and B$_2$S \cite{Zhao2018}. 
These predictions were followed by a number of studies predicting structurally similar anisotropic Dirac materials \cite{C8RA08291J,doi:10.1021/acs.jpcc.0c00696}.
In 2021, Bykov \emph{et al.} \cite{Bykov2021} have reported a high pressure synthesis of a triclinic phase of beryllium tetranitride BeN$_4$. Upon decompression to ambient conditions, BeN$_4$ transforms into a compound with atomic-thick BeN$_4$ layers coupled be weak van der Waals bonds. Theoretical calculations demonstrated that the electronic lattice of each BeN$_4$ layer (beryllonitrene) can be described by a distorted honeycomb lattice, resulting in anisotropic Dirac cones with a large anisotropy ratio of $v_y/v_x \sim 2.5$. The discovery of BeN$_4$ has stimulated the prediction of other isostructural 2D metal-nitrides \cite{Mortazavi2021}, some of which (e.g., MgN$_4$) exhibit similar anisotropic properties. 
Also, it is worth mentioning a wide range of 2D designer materials composed of molecular precursors, some of which are predicted to host anisotropic Dirac fermions \cite{Udo2021}.

\begin{figure}[t]
\centering
\mbox{
\includegraphics[width=0.8\linewidth]{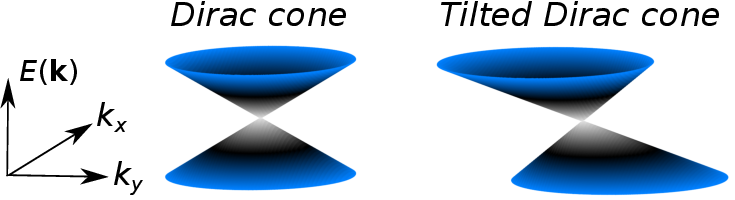}
}
\caption{Energy dispersion of a normal anisotropic Dirac cone [$v_t=0$ in Eq.~(\ref{E_tilt})] and a tilted anisotropic Dirac cone [$v_t\neq0$].}
\label{fig3}
\end{figure}

Among the 2D anisotropic Dirac semimetals, another special class of materials can be distinguished, namely, materials with \emph{tilted} anisotropic Dirac cones. Around the crossing points, the corresponding Hamiltonian can be written as
\begin{equation}
    H = \hbar v_x\sigma_xq_x + \hbar v_y\sigma_yq_y + \hbar v_t\sigma_0 q_y,
    \label{H-tilt}
\end{equation}
where $\sigma_x$, $\sigma_y$, and $\sigma_0$ are the (2$\times$2) Pauli matrices, $\sigma_0$ being the unit matrix, and $v_t$ is a tilting parameter. The Hamiltonian (\ref{H-tilt}) differs from the Hamiltonian  (\ref{H-anisD}) by the last term, which induces a tilting of the Dirac cone in a specific direction ($y$ in our case). The corresponding energy dispersion is given by
\begin{equation}
    E({\bf q}) = \hbar v_tq_y \pm \sqrt{\hbar^2 v^2_xq^2_x + \hbar^2 v_y^2q_y^2},
    \label{E_tilt}
\end{equation}
which is schematically shown in Fig.~\ref{fig3} for the cases $v_t\neq0$ and $v_t=0$.
One can see that the tilt leads to an elongation of the Fermi contour along the $y$ direction. In the purely tilted case ($v_x=v_y$), the Fermi velocities remain anisotropic with the ratio $\tilde{v}_x/\tilde{v}_y = \sqrt{1-\gamma^2}$, where $\gamma=v_t/v_y$, and in the majority of practically-relevant cases $\gamma<1$.
The averaged Fermi velocity is renormalized by the tilting, preserving the linear dependence of DOS \cite{Herrera2019}, 
\begin{equation}
    g(E) = \frac{|E|}{2\pi \hbar^2 v_x v_y}(1-\gamma^2)^{-3/2}.
\end{equation}
As a consequence, apart from a renormalization of the Fermi velocity the presence of tilting does not change the behavior in magnetic fields, $E_n\propto\sqrt{nB}$, but only affects spacing between the Landau levels \cite{Goerbig2008}.

A physical example of the system with tilted Dirac cones includes the so-called quinoid-type lattice deformation of the honeycomb lattice with nearest- and next-nearest-neighbor interactions \cite{Goerbig2008}. In this type of deformation, the lattice is deformed in a way, in which only the bonds parallel to the deformation axis are modified. In practice, it is hardly possible to maintain this condition, making such deformations difficult to achieve by mechanical means. Tilted Dirac cones were predicted to appear in partially hydrogenated graphene with the chemical composition C$_6$H$_2$ \cite{Lu2016}, as well as in more exotic structures like orthorhombic phosphorus nitride \cite{Nie2020}. Experimentally, there are indirect evidences of the tilted Dirac cone formation in the organic semiconductor $\alpha$-(BEDT-TTF)$_2$I$_3$ under pressure \cite{Katayama2006,Hirata2016,Hirata2017}. 
Another prominent example of 2D semimetals with tilted Dirac cones is the 8-$Pmmn$ phase of borophene \cite{Oganov2014,Lopez2016}. Despite multiple experimental realizations of 2D borophene \cite{Mannix2015,Feng2016},
confirming the existence of 2D anisotropic Dirac cones observed by ARPES \cite{Baojie2017,Feng2018}, the experimentally available phases $\beta_{12}$ and $\chi_3$ turn out to be metallic, which limits their application in the context of massless Dirac fermions. Recently, a number of Pt-based 2D Dirac semimetals which combine tilted cones and strong spin-orbit coupling have been predicted \cite{Dong2022}. Such combination gives rise to the possibility to realize 2D quantum spin Hall insulators with orientation-dependent properties.

It is worth noting that chiral properties of the Hamiltonian (\ref{H-tilt}) give rise to unique physics, closely related to that of Weyl-Dirac topological systems. Next to that, materials with tilted Dirac cones constitute a promising platform for probing the effects of strong electron correlations, including excitonic pairing and anomalous spin fluctuations \cite{Hirata2017,Ohki2020}.

\subsection{\label{sec2d}Semi-Dirac semimetals}

Let us now put $|t_2|=2|t_1|$ in the Hamiltonian (\ref{H_expand}), which corresponds to the transition point between the gapped state ($|t_2|>2|t_1|$) and the anisotropic Dirac state ($|t_2|<2|t_1|$). In this special case, we have a gapless dispersion
\begin{equation}
    E({\bf q})=\pm \sqrt{ \left( \frac{\hbar^2 q_x^2}{2m_x} + \frac{\hbar^2 q_y^2}{2m_y}\right)^2 +      \hbar^2 v_x^2q_x^2 }.
    \label{Eq_semi}
\end{equation}
Since there is no gap, in the limit ${\bf q}\rightarrow0$ we have a linear dependence in the $x$-direction, and a quadratic one in the $y$-direction [Fig.~\ref{fig1}(f)], which is referred to as the semi-Dirac dispersion. In this case, DOS acquires a square-root dependence \cite{Montambaux2009-2},
\begin{equation}
    g(E) = C\frac{\sqrt{m_y}}{\hbar^2 v_x}\sqrt{|E|},
\end{equation}
where $C=K(2^{-1/2})/\pi^2 \simeq 0.188$.
One of the most unusual consequence of the semi-Dirac energy dispersion is the behavior in magnetic fields.
The energies of the Landau levels vary with the field and the level number as \cite{Goerbig2008,Montambaux2009-2}
\begin{equation}
    E_n = \pm A \hbar(m_yv_x^2)^{1/3} [(n+1/2)\omega_c]^{2/3},
\end{equation}
where $\omega_c=eB/m_y$ and $A=\sqrt{C/2^{2/3}}\simeq 1.173$. Therefore, we see that $E_n \propto [(n+1/2)B]^{2/3}$, which is essentially different from the behavior of purely quadratic ($\propto B$) or purely linear ($\propto \sqrt{B}$) bands. The $B^{2/3}$ scaling of Landau levels has been recently observed experimentally in the context of a prototypical nodal-line semimetal ZrSiS, where the semi-Dirac spectrum is realized around some of the crossing points \cite{shao2023semidirac}.

The unusual Landau level scaling leads to a modification of the orbital susceptibility, compared to the Dirac systems. For a honeycomb lattice with nearest-neighbor interactions, the temperature-dependence of the orbital susceptibility at zero doping can be written as $\chi(T) = \chi_{\mathrm{pl}} - 3\chi_0/(16\pi T)$ \cite{Stauber2011}, where $\chi_0 = \mu_0\hbar^{-2}e^2|t|a^2$, and $\chi_{\mathrm{pl}}\approx0.089\chi_0$ is the so-called paramagnetic plateau, originating from the interband coupling. For semi-Dirac fermions, the susceptibility also shows a diamagnetic divergence at low temperatures, but with an essentially weaker dependence, $\chi(T)\propto 1/\sqrt{T}$ \cite{Raoux2015}. It should be noted that this dependence arises from the effects of interband coupling, and cannot be described within the famous Landau-Peierls formula \cite{Vonsovskii_book}, which predicts an $\propto1/T$ dependence. The energy dependence of the susceptibility also turns out to be different for semi-Dirac systems, yielding a $\propto-1/\sqrt{|E|}$ scaling \cite{Raoux2015,Pickett2012}, whereas for Dirac fermions one has $\chi(T=0) \approx \chi_{\mathrm{pl}}$ at finite doping. Generally, semi-Dirac systems are expected to have diamagnetic orbital response in a wider range of temperatures and doping levels.

Before considering realistic examples, we first note that the spectrum of the semi-Dirac form (\ref{Eq_semi}) has appeared even before the discovery of graphene, in the context of a topologically nontrivial phase (``A'' phase) of superfluid helium-3 \cite{Volovik2001}.
In the material science context, the existence of the semi-Dirac point has been predicted from first-principles calculations in the half-metallic VO$_2$-TiO$_2$ heterostructures, where three or four VO$_2$ layers are confined within TiO$_2$ \cite{Pardo2009,Pardo2010}. In Ref.~\onlinecite{Banerjee2009}, it was demonstrated that the energy dispersion in the form of Eq.~(\ref{Eq_semi}) can be obtained from the 3-band TB model of spinless fermions describing the VO$_2$ trilayer. Later, it was realized that there is another type of the semi-Dirac metal, described by the Hamiltonian
\begin{equation}
H =
    \begin{pmatrix}
        0 & \frac{\hbar^2q_y^2}{2m_y} - \hbar v_xq_x -i\hbar \alpha q_xq_y \\
        \frac{\hbar^2 q_y^2}{2m_y} - \hbar v_xq_x +i\hbar \alpha q_xq_y & 0 
        \end{pmatrix},
    \label{H_type2}
\end{equation}
which is different from Eq.~(\ref{tb-honey}), but resulting in the energy dispersion that satisfies the semi-Dirac character, i.e. yields linear and quadratic dispersion along two perpendicular directions. It was shown that this particular Hamiltonian describes the half-metallic superlattice (TiO$_2$)$_5$/(VO$_2$)$_3$ at low energies. At the same time, the broken time-reversal symmetry results in remarkable topological properties of the model, namely, the emergence of a Chern-insulating state \cite{Huang2015}. Indeed, if spin-orbit coupling is included in the Hamiltonian (\ref{H_type2}) in the form $H\rightarrow H + m_z \sigma_z$, it opens a gap at the semi-Dirac point, and gives rise to a nontrivial topology with the Chern number ${\cal C}=-2$. In turn, the nonzero Chern number results in the anomalous Hall effect with a quantized Hall plateau at $\sigma_{xy}=-{\cal C}e^2/h$, which is reproduced by first-principles calculations \cite{Huang2015}.

\subsection{\label{sec2e}Flat band systems}
In recent years, one can see a growing interest  to the so-called flat-band materials, exhibiting anomalies in the band structure that give rise to the van Hove singularities in DOS, closely related to the correlation effects and various forms of many-body instabilities. In the context of 2D materials, this interest was connected to the flat band formation in twisted bilayer or multilayer van der Waals heterostructures, an important section of the novel field called twistronics \cite{Carr2017}. This interest was strongly enhanced by the discovery of unconventional superconductivity in ``magic angle'' twisted bilayer graphene \cite{Cao2018,Yankowitz2019,Andrei2020}. This is a huge field by itself, with enormous amount of literature, and we have to avoid its discussion here focusing on other aspects of flat band formation in 2D materials. 

Flat bands are known to exist in 2D materials (e.g., InSe) as a result of band structure peculiarities, such as saddle points in the vicinity of the band edges \cite{Lugovskoi2019}. Such peculiarities can be realized, e.g., on a triangular lattice model with nearest- and next-nearest hoppings opposite in sign. Upon doping, such systems indeed demonstrate unconventional behavior, for instance, strong electron-phonon coupling or charge density wave formation \cite{Stepanov2022}. The energy dispersion that gives rise to flat bands should include high-order terms. A minimum model can be written as
\begin{equation}
    E({\bf q}) = \frac{1}{2}\hbar^2\alpha q_x^2 - \frac{1}{4}\hbar^4\beta q_x^4 - \frac{1}{2}\hbar^2\alpha'q_y^2,
\label{vhs}
\end{equation}
which has the form of a saddle with the highest band edge at $E_{\mathrm{max}}=\alpha^2/4\beta$. Around $E=0$, which is the saddle point, DOS diverges as $\propto \mathrm{ln}(E_\mathrm{max}/E)/\sqrt{\alpha \alpha'}$ \cite{Ziletti2015}. However, if the coefficient at $q_x^2$ were zero, we would have a much stronger singularity in DOS $\propto E^{-1/4}$. In this case, as well as in the situation when the $q_x^4$ term dominates, one expects a superlinear dependence of the Landau levels, $E_n \propto [(n+\gamma)B]^{4/3}$ \cite{Goerbig2008}. The dispersion of the form (\ref{vhs}) was assigned to the valence band in monolayer BP under compressive strain, followed by the prediction of a ferromagnetic instability in this system upon hole doping \cite{Ziegler2017}.

Recently, an extremely anisotropic van der Waals layered system, CrSBr, has been discovered. CrSBr is an orthorhombic semiconductor with an energy gap of 1.5--2.0 eV \cite{Bianchi2023}, composed of ferromagnetic layers coupled antiferromagnetically.
Apart from interesting magnetic properties, offering diverse opportunities for their tunability \cite{Klein2022,Rudenko2023}, 
CrSBr features flat bands in the conduction band bottom, giving rise to an effective electron mass ratio of $m_x/m_y \sim 50$, behaving like a quasi-1D material \cite{Wu2022,Klein2023}.

Flat bands can be also formed by specific correlation effects arising when 2D Van Hove singularity lies close enough to the Fermi energy \cite{Irkhin2002,Yudin2014}. This is accompanied by pinning of the singularity to the Fermi energy in a relative broad interval of doping \cite{Irkhin2002}. 
Among anisotropic 2D materials, BeN$_4$ seems to be a potential candidate for practical realization of these effects due to relative proximity of the van Hove singularities to the Fermi energy \cite{Bykov2021}.

\section{\label{sec3}Dielectric screening and optical response}
\subsection{\label{sec3a}Optical conductivity}
For anisotropic systems, the optical conductivity is given by a tensor with unequal diagonal components. The symmetric off-diagonal components depend on the reference frame and can be nullified by an appropriate choice of the principal axes. In what follows, we neglect antisymmetric off-diagonal components of this tensor, i.e., ignore gyrotropic effects that require time-reversal symmetry breaking. Therefore, for the cases under consideration, we restrict ourselves the two diagonal components: $\sigma_{xx}(\omega)$ and $\sigma_{yy}(\omega)$.

The $xx$-component of the optical conductivity of anisotropic 2DEG with the dispersion given by Eq.~(\ref{el-gas}) can be written within the Drude model as
\begin{equation}
\sigma_{xx}(\omega) = \frac{\sigma_{xx}^0}{1-i\omega /\eta}
\label{drude},
\end{equation}
where $\sigma_{xx}^{0} = ne^2/m_x\eta$ is the dc conductivity, $\eta$ is the scattering rate, and 
\begin{equation}
n = \frac{\sqrt{m_xm_y}}{2\pi \hbar^2}T\,\mathrm{ln}\left[ \mathrm{exp}(\mu/T)+1 \right]
\end{equation} is the charge density determined by the chemical potential $\mu$. At $T=0$, $n = \mu\sqrt{m_xm_y}/2\pi\hbar^2$ and we can recast the dc conductivity as
\begin{equation}
    \sigma^0_{xx} = \frac{e^2}{2\pi \hbar^2 \eta}\sqrt{\frac{m_y}{m_x}}\mu,
\end{equation}
which is given per spin per valley. A symmetric expression is obtained for $\sigma^{0}_{yy}$ by the replacement $x\rightleftarrows y$. Unlike the isotropic Drude model, the components of the conductivity in the anisotropic model acquire dependence on the effective mass ratio $m_y/m_x$, while the average quantity $\sqrt{\sigma^0_{xx}\sigma^0_{yy}}$ remains independent and coincides with the case $m_x=m_y$.

The interband contribution to the conductivity can be calculated from the Kubo formula as a response to an electric field with frequency $\omega$ and polarization along the $i$ direction. The real (absorptive) part of the conductivity can be written as \cite{Green-Book}
\begin{eqnarray}
\notag
    \sigma_{ii}(\omega) = \pi\frac{e^2}{\hbar}\frac{1}{\omega}\int d\omega [f(\varepsilon)- f(\varepsilon+\omega)] \\
    \times \int \frac{d{\bf k}}{4\pi^2}\, \mathrm{Tr}[v_i \, A({\bf k,\varepsilon}) \, v_i \, A({\bf k,\varepsilon+\omega})], 
\label{kubo}
\end{eqnarray}
where $v_i=-\frac{i}{\hbar}[H,r_i]$ is the $i$-component of the velocity operator with $r_i$ being the position operator, $f(\varepsilon)=\{1+\mathrm{exp}[(\varepsilon - \mu)/k_BT]\}^{-1}$ is the Fermi-Dirac distribution function, and $A({\bf k},\varepsilon)=-(1/\pi)\mathrm{Im}[G({\bf k},\varepsilon)]$ is the spectral function evaluated from the Green's function $G({\bf k},\varepsilon)=[(\varepsilon + i\eta)I - H]^{-1}$.

\begin{figure}[t]
\centering
\mbox{
\includegraphics[width=0.8\linewidth]{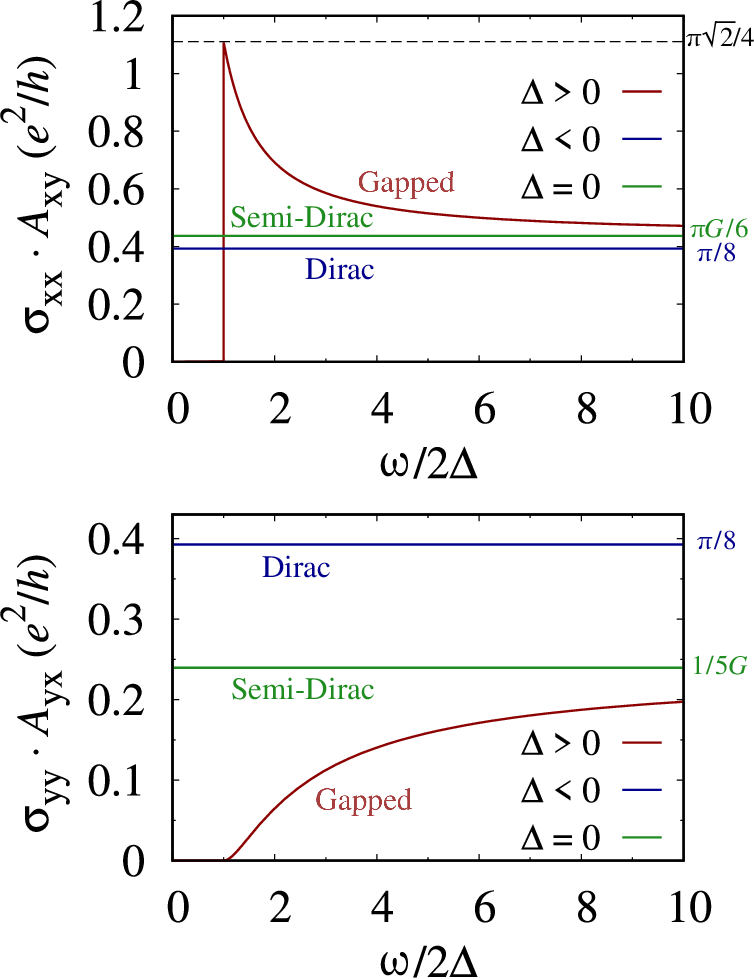}
}
\caption{Interband contribution to the optical conductivity
calculated along the $x$- and $y$- directions for anisotropic gapped semiconductor ($\Delta>0$), anisotropic Dirac semimetal ($\Delta<0$) and semi-Dirac semimetal ($\Delta=0$) using the Kubo formula at $T=0$ and $\eta=0$. The conductivity is given per spin per valley with respect to the dimensionless anisotropy parameter $A_{xy}=A^{-1}_{yx}$, which reads $A_{xy}=v_y/v_x$ (Dirac case) and $A_{xy}=\sqrt{\omega/m_yv_x^2}$ (other cases).
}
\label{fig4_conductivity}
\end{figure}

Let us now consider anisotropic semiconductor with the Hamiltonian (\ref{H_expand}), $\Delta>0$ and $\mu=0$. In the limit $v_x^2/\Delta \ll m_x^{-1}$, i.e., when the linear term can be neglected, the velocity operator is diagonal in the basis of eigenstates of the Hamiltonian, ensuring vanishing interband conductivity. In the opposite limit $v_x^2/\Delta \gg m_x^{-1}$ we deal with the hybrid dispersion yielding non-vanishing off-diagonal components of the velocity operator. In this case, the conductivity at $T=0$ and $\eta=0$ can be written as
\begin{eqnarray}
\label{sigma_xx-gap}
    \sigma^{\mathrm{inter}}_{xx}(\omega) = 
    \frac{e^2}{h}\frac{\pi G}{6} \sqrt{\frac{m_yv_x^2}{\omega}} f_x\left(\frac{2\Delta}{\omega}\right), \\
    \label{sigma_yy-gap}
    \sigma^{\mathrm{inter}}_{yy}(\omega) = 
    \frac{e^2}{h}\frac{1}{5G} \sqrt{\frac{\omega}{m_yv_x^2}} f_y\left(\frac{2\Delta}{\omega}\right).
\end{eqnarray}
where $G=\Gamma(\frac{1}{4})^2/2\sqrt{2\pi^2} \approx 0.835$ is the Gauss's constant, and $f_x$, $f_y$ are monotonic functions such that $f_x(0)=1$, $f_x(1)=3/\sqrt{2}G$, $f_y(0)=1$, and $f_y(1)=0$. The quantity $A_{xy}=\sqrt{\omega/m_yv_x^2}$ plays the role of a frequency-dependent anisotropy parameter. In Fig.~\ref{fig4_conductivity}, we show the $xx$ and $yy$ components of the conductivity calculated as a function of frequency. The intraband conductivity has an onset at the band gap $\omega=2\Delta$, displaying drastically different behavior along the $x$ and $y$ directions. Near the onset, one has $\sigma_{xx} \sim \omega^{-3/2}$ and $\sigma_{yy} \sim \omega^{3/2}$. It is important to note that $\sigma_{xx}$ does not diverge at $\omega=2\Delta$, reaching its maximum $\sigma^\mathrm{max}_{xx}=\frac{e^2}{h}\frac{\pi}{4} \sqrt{m_yv_x^2/\Delta}$. This result is different from the result of Carbotte \emph{et al.}~\cite{Carbotte2019}, who obtained $\sigma_{xx} \sim [\omega^2 - 4\Delta^2]^{-1/4}$ considering a slightly different model, where the energy gap is introduced from the sublattice symmetry breaking, resulting in the appearance of diagonal terms in the Hamiltonian.
Interestingly, unlike the anisotropic Drude model discussed earlier, the quantity $\sqrt{\sigma_{xx}\sigma_{yy}}$ is now independent of the material's constants $v_x$ and $m_y$, and only depends on $\Delta$. This dependence, however, disappears in the limit $\omega \gg \Delta$, yielding $\sqrt{\sigma_{xx}\sigma_{yy}} = \frac{e^2}{h}\sqrt{\frac{\pi}{30}}$, which is approximately 20\% smaller than the universal optical conductivity of graphene per spin per valley $\sigma=\frac{e^2}{h}\frac{\pi}{8}$ \cite{Katsnelson-Book}.

Despite being based on a simple model, Eqs.~(\ref{sigma_xx-gap}) and (\ref{sigma_yy-gap}) describe the conductivity of anisotropic semiconductors remarkably good. In Ref.~\onlinecite{Rudenko2015}, optical conductivity was calculated in few-layer and bulk BP along the armchair ($xx$) and zigzag ($yy$) directions based on a TB model. 
The TB calculations reproduce the characteristic behavior of the conductivity along the $x$ and $y$ directions predicted by Eqs.~(\ref{sigma_xx-gap}) and (\ref{sigma_yy-gap}) not only for monolayer, but also for multilayer BP. In the limit of bulk BP, the peak at $\omega=2\Delta$ disappears, indicating that this peculiar behavior of the anisotropic conductivity is the effect of dimensionality. In 2D BP, $\sigma_{xx}\gg \sigma_{yy}$ for all relevant frequencies, meaning that the anisotropy parameter $A_{xy}\gg 1$. This, in turn, highlights the importance of the linear term in the Hamiltonian of BP-like semiconductors, which plays a key role in the strongly anisotropic optical response.
A huge difference between the $xx$ and $yy$ components of the optical conductivity indicates that light polarized along the $x$ is absorbed, while light polarized along the $y$ direction is transmitted.  Such unique optical response make anisotropic materials potentially applicable for optical polarizers.

We now proceed to the optical conductivity of anisotropic Dirac materials. Optical properties of anisotropic Dirac fermions were analyzed in the context of strained graphene \cite{Pereira2010,Pellegrino2010,OlivaLeyva2014} and anisotropic honeycomb lattices \cite{OlivaLeyva2016,OlivaLeyva2017}. 
Within the Dirac cone approximation given by the Hamiltonian (\ref{H-anisD}), the $xx$ components of the optical conductivity tensor at finite temperature in the presence of random scattering and zero doping can be written (per spin per valley) as \cite{Herrera2019} 
\begin{equation}
\sigma^{\mathrm{inter}}_{xx}(\omega) = \frac{\pi e^2}{8h}\frac{v_x}{v_y}\left\{ \mathrm{tanh} \, \Omega  + \frac{1}{2\pi} \frac{\beta \eta}{\Omega^2} \, \mathrm{ln} \,{\mathrm{cosh}{(2\Omega})} \right\},   
\label{sigmaxx_t}
\end{equation}
where we denote $\Omega=\frac{\beta\omega}{4}$ and
$\beta=1/k_BT$. The $yy$ component is obtained by the replacement $x \rightleftarrows y$. At zero temperature, this expression simplifies to 
\begin{equation}
    \sigma^{\mathrm{inter}}_{xx}(\omega) = \frac{v_x}{v_y}\frac{\pi e^2}{8h}\left(  1 + \frac{4\eta}{\pi \omega}   \right).
    \label{sigmaxx_t0}
\end{equation}
For the isotropic dispersion $v_x=v_y$ and $\eta=0$, Eq.~(\ref{sigmaxx_t0}) reduces to the universal conductivity of graphene per spin per valley, $\pi e^2/8h$. Therefore, the main effect of anisotropy on the optical conductivity of Dirac fermions is the appearance of the $v_x/v_y$ scaling factor, provided $\eta$ is a direction-independent constant. The expressions (\ref{sigmaxx_t}) and (\ref{sigmaxx_t0}) are only valid if $\omega \ll 2\Delta$, where $2\Delta=4t_1+2t_2$ is the inverse gap of the distorted honeycomb lattice, which corresponds to the energy separation between the van Hove singularities in DOS. While for graphene $2\Delta \simeq 6$~eV, it can be considerably smaller in anisotropic Dirac materials. For instance, in monolayer BeN$_4$ one has $2\Delta \simeq 1$ eV \cite{Bykov2021}, limiting practical application of the equations given above. In the situation when the inverse gap is small, i.e. when two Dirac points are close to each other, the conductivity becomes $\omega$-dependent even in the clean limit ($\eta=0$). A transition of the optical conductivity from the isotropic limit ($t_2=t_1$) to the merging point ($t_2=2t_1$) has been analyzed by Ziegler and Sinner \cite{Ziegler2017}.

When the anisotropy originates purely from tilting of the Dirac cone ($v_x=v_y$), one has $\sigma_{yy}/\sigma_{xx}=\sqrt{1-\gamma^2}$ \cite{Herrera2019}. Optical conductivity of the tilted Dirac semimetals with the Hamiltonian (\ref{H-tilt}) in the most general case with $v_x\neq v_y$ and $v_t\neq 0$ has been analyzed in detail in Refs.~\onlinecite{Herrera2019,Mojarro2021}.

As in the case of graphene \cite{Katsnelson-Book}, optical conductivity of undoped anisotropic Dirac semimetals within the Dirac cone approximation is purely real. Finite doping gives rise to the emergence of the imaginary part of the conductivity, which can be restored from the expressions above using the Kramers-Kroning relations. Besides, finite doping in Dirac semimetals ensures a nonzero contribution to the intraband conductivity. The scaling behavior of the interband conductivity in anisotropic Dirac semimetals is generalized to the intraband (Drude) one, yielding 
\begin{equation}
\sigma^{\mathrm{intra}}_{xx}(\omega) = \frac{v_x}{v_y}\frac{\pi e^2}{2h} \mu \, \delta(\omega). 
\end{equation}

Optical conductivity of semi-Dirac systems has been studied in Ref.~\onlinecite{Ziegler2017} using tight-binding calculations on the distorted honeycomb lattice, for which $t_2=2t_1$. At charge neutrality ($\mu=0$), a strong dependence of the anisotropy at small frequencies with $\sigma_{xx}\sim \omega^{-1/2}$ and $\sigma_{yy}\sim \omega^{1/2}$ was found. This behavior is in line with Eqs.~(\ref{sigma_xx-gap}) and (\ref{sigma_xx-gap}). Setting $\Delta =0$, we obtain in the limit of small $\omega$ the result of Carbotte \emph{et al.} \cite{Carbotte2019}, namely, 
\begin{equation}
    \sigma^{\mathrm{inter}}_{xx} = \frac{e^2}{h}\sqrt{\frac{m_yv_x^2}{\omega}}\frac{\pi G}{6} \text{, } \quad    \sigma^{\mathrm{inter}}_{yy} = \frac{e^2}{h}\sqrt{\frac{\omega}{m_yv_x^2}}\frac{1}{5G}.
\end{equation}
Again, the averaged conductivity $\sqrt{\sigma_{xx}\sigma_{yy}}=\frac{e^2}{h}\sqrt{\frac{\pi}{30}}$ is a material-independent constant. 
The intraband contribution in the semi-Dirac case 
scales as $\sigma^{\mathrm{intra}}_{xx} \sim \mu^{1/2}$ and $\sigma^{\mathrm{intra}}_{yy} \sim \mu^{3/2}$ \cite{Carbotte2019}, which differs from the result of the Drude model.
Complex optical conductivity and its dependence on the field orientation for a semi-Dirac system has been studied by Sanderson \emph{et al.} \cite{Sanderson2018}. The imaginary part of the conductivity has been shown to have a peculiar behavior. In particular, depending on the orientation angle and frequency, there is a region in which the imaginary part of the transverse conductivity is greater than the longitudinal one. Moreover, one can find a regime with purely transverse complex conductivity.

\subsection{\label{plasmons}Dynamical polarizability and plasmons}
Within linear response theory, the polarizability can be obtained as a charge density response to an external scalar potential, $\delta n_{\omega {\bf q}}=\Pi(\omega, {\bf q})\delta \Phi_{\omega {\bf q}}$. In the basis of the energy states, the polarizability matrix of a 2D multiband system can be written as \cite{vignale_book,vk_book,Katsnelson-Book}
\begin{equation}
    \label{polariz_general}
    \Pi(\omega,{\bf q})= \int \frac{d{\bf k}}{(2\pi)^2}\frac{|\langle \psi_{m{\bf k}} | \psi_{n{{\bf k} + {\bf q}}} \rangle|^2  \, (f_{m\bf k} - f_{n{\bf k}+{\bf q}}) }{E_{{m}{\bf k}} - E_{{n}{\bf k}+{\bf q}} + \omega + i\eta},
\end{equation}
where $E_{m{\bf k}}$ and $\psi_{m{\bf k}}$ are the eigenvalues and eigenvectors of the electronic Hamiltonian, and $f_{m{\bf k}}$ is the Fermi occupation factor.

Analytical expression for the dynamical polarizability of anisotropic 2DEG at $T=0$ was
obtained by Rodin and Castro Neto \cite{Rodin2015}. In the pure limit ($\eta \rightarrow 0$), it reads 
\begin{equation}
    \frac{\Pi(\omega,{\bf q})}{g(\mu)} = \frac{1}{2}\sum_{j=\pm1}\left[ \prod_{l=\pm1} \sqrt{1 + j\frac{\hbar\omega}{E({\bf q})} + l\frac{2\sqrt{\mu}}{\sqrt{E({\bf q})}}} -1 \right],
    \label{P-rodin}
\end{equation}
where $\mu$ is the chemical potential $E({\bf q})$ is the dispersion of anisotropic 2DEG [see Eq.~(\ref{el-gas})], and $g(\mu)$ 
is the corresponding DOS. In the static limit ($\omega \rightarrow 0$), Eq.~(\ref{P-rodin}) reduces to the polarization function that was first obtained by Low \emph{et al.} in the context of monolayer BP \cite{Low2014},

\begin{equation}
\Pi({\bf q}) = g(\mu) \left[ \sqrt{1 - \frac{8\mu/\hbar^2}{q_x^2/m_x + q_y^2/m_y}} -1 \right ].    
\label{polariz_2d}
\end{equation}
Equation (\ref{polariz_2d}) is valid for $q\ge2|{\bf k}_\mathrm{F}|$, where ${\bf k}_\mathrm{F}$ is the direction-dependent Fermi wave vector, and the expression under the square root is not negative. In the opposite case, $q\le2|{\bf k}_\mathrm{F}|$, the polarization reduces to a constant $\Pi({\bf q})=g(\mu)$, like in the isotropic case. Therefore, the scattering processes involving small momenta in anisotropic 2DEG is nonetheless isotropic. In this regime, the screening is metallic and the dielectric function is well described by the Thomas-Fermi approximation, yielding $\varepsilon({\bf q})= 1+ \kappa/|{\bf q}|$, where $\kappa=2\pi e^2 g(\mu)$ is the screening wave vector. However, this result is only applicable in the absence of disorder and at $T=0$. With nonzero temperature and disorder, the static screening is, strictly speaking, anisotropic \cite{Low2014}.

\begin{figure}[t]
\centering
\mbox{
\includegraphics[width=1.0\linewidth]{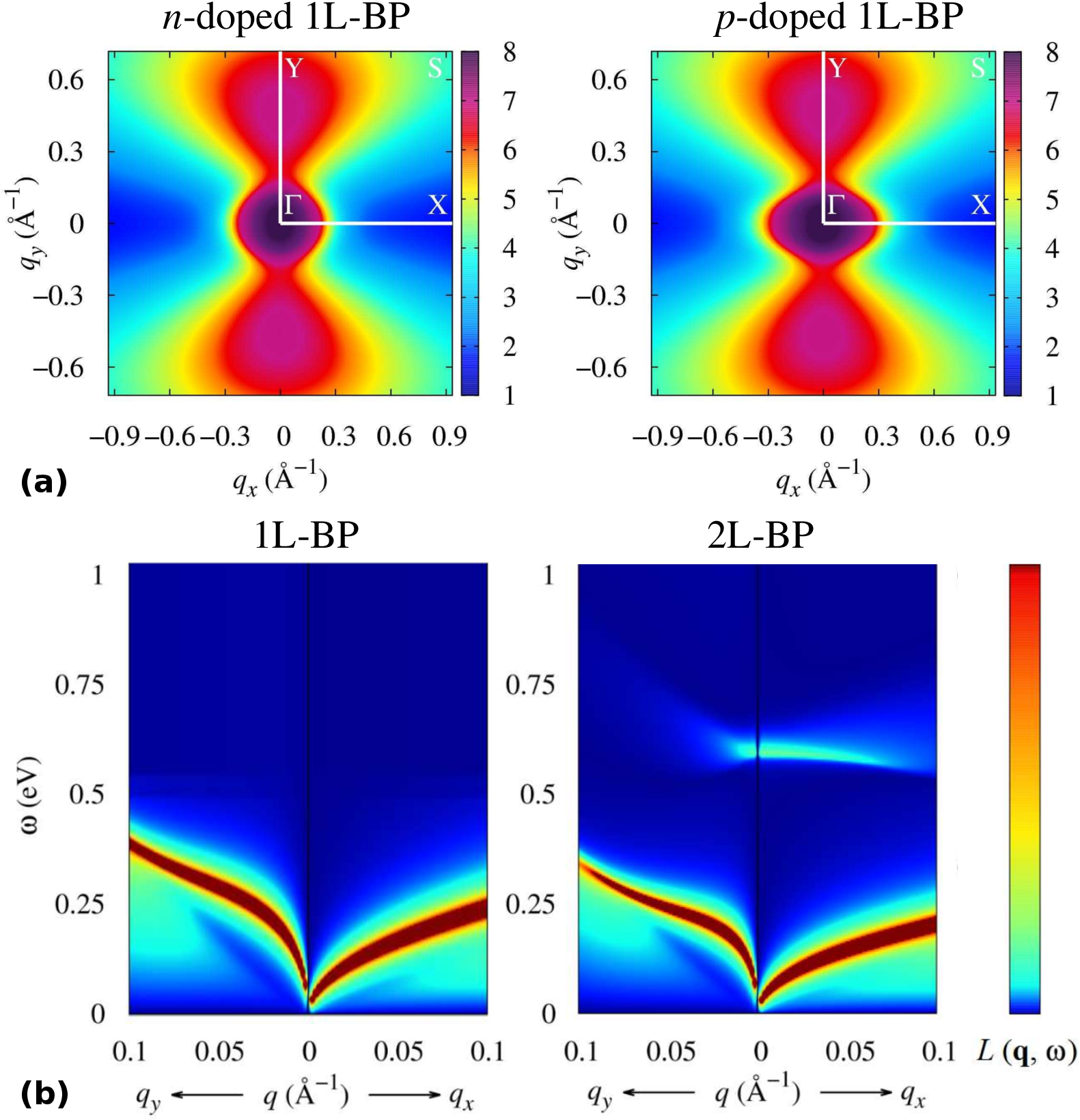}
}
\caption{(a) Static dielectric function $\varepsilon({\bf q})$ resolved over the Brillouin zone calculated within RPA for electron-doped (left) and hole-doped (right) monolayer BP using a TB model \cite{Prishchenko2017}. In accordance with Fig.~\ref{fig2}, $x$ and $y$ correspond to the heavy mass (zigzag) and light mass (armchair) directions, respectively. 
(b) Electron-loss function $L({\bf q},\omega)=-\mathrm{Im}[1/\varepsilon({\bf q,\omega})]$ calculated using the same method for electron-doped single-layer (left) and bilayer-BP (right), demonstrating anisotropy of the plasmon dispersion. 
In all cases, doping corresponds to the carrier concentration of $10^{13}$ cm$^{-2}$.
}
\label{epsilon_eels}
\end{figure}

Figure~\ref{epsilon_eels}(a) shows the static dielectric function calculated within the random phase approximation (RPA) \cite{vk_book, vignale_book}, 
\begin{equation}
\varepsilon(\omega,{\bf q})=1-v({\bf q})\Pi(\omega,{\bf q}),
\label{rpa}
\end{equation} 
for electron- and hole-doped monolayer BP at $T=300$~K \cite{Prishchenko2017}. The calculations were performed based on a four-site TB model \cite{Rudenko2015} with the Coulomb interaction $v({\bf q})$ determined from first principles, i.e. beyond the long-wavelength limit. As can be seen, at small ${\bf q}$ the dielectric function is indeed essentially isotropic. Beyond this region, however, $\varepsilon({\bf q})$ becomes highly anisotropic, which is determined by the anisotropy of the polarization function, not the Coulomb interaction. Interestingly, for $q \ge 2|{\bf k}_\mathrm{F}|$, the screening along the heavy mass direction ($x$) is strongly suppressed, $\varepsilon(q_x,0) \sim 1$. This behavior can be understood keeping in mind 
that transitions with the momentum transfer along $x$ have less weight, thus providing smaller contribution to the polarization. At the Brillouin zone edges, the screening anisotropy reaches its maximum, yielding $\varepsilon(Y) / \varepsilon(X) \approx 4$.

Anisotropic screening of BP surface has been experimentally demonstrated by means of scanning tunneling microscopy (STM) \cite{Kiraly2019}. The authors of Ref.~\onlinecite{Kiraly2019} deposited potassium adatoms on the BP surface and observed the development of 1D potassium structures. Remarkably, these structures orient along the armchair direction, i.e., orthogonal to the direction along which the diffusion barrier is lower, demonstrating that the ordering is not driven by diffusion energy barriers.
The formation of potassium chain-like structures is ascribed to the anisotropic screening of 2DEG formed near the BP surface, which affects the Coulomb interaction between the potassium adatoms, governing direction-dependent interatomic spacing. The observed ordering was further interpreted in terms of Friedel oscillations confirmed by coverage-dependent measurements, shown to be dependent on the orientation, in agreement with theory.  

Anisotropic interactions between adatoms on anisotropic surfaces may be utilized for practical applications. Some of the transition metal adatoms supported on BP, such as Co and Fe, demonstrate very special property of valence bistability, giving rise to the realization of single-atom orbital memory \cite{Kiraly2018,Knol2022,Kiraly2022}. Taking advantage of strong anisotropy of the adatom--adatom interactions at the surface of BP, orbital bistability of Co adatoms was exploited in the demonstration of opportunities for neuromorphic computations \cite{Kiraly2021}.

Let us now discuss the role of anisotropy effects in the context of dynamical screening.
The presence of dynamically screened Coulomb interaction in an electron gas gives rise to the emergence of self-sustained longitudinal oscillations of the charge density, or plasmons \cite{vk_book, vignale_book}. The properties of plasmon excitations in anisotropic materials turn out to be essentially different from their isotropic analogs. Apart from direction-dependent dispersion relation of plasmons, anisotropy may results in qualitatively new effects. One of them is the direction-dependent Landau damping with the possibility of a situation when the plasmon is undamped in one direction and damped in another direction \cite{DasSarma2021}. Also, anisotropic materials allow for a regime of the so-called hyperbolic plasmons, highly unusual excitations with very large wave vectors and large photonic density of states inside the material \cite{Nemilentsau2016}.

In the long-wavelength limit, the plasmon oscillations are well described by classical Maxwell theory. The dispersion relation of a bound mode at the surface of a homogeneous dielectric medium can be described by the equation \cite{Nemilentsau2016} (in cgs units)
\begin{equation}
  \frac{q^2_x}{\sigma^{''}_{yy}} + \frac{q^2_y}{\sigma^{''}_{xx}} = \frac{|{\bf q}|\omega}{2\pi} \left(\frac{1}{\sigma^{''}_{xx}\sigma^{''}_{yy}} -\frac{\pi}{c^2}  \right) ,
  \label{maxwell}
 \end{equation}
which assumes $|{\bf q}|\gg \omega/c$ with $c$ being the speed of light, and $\omega/\eta \gg 1$, i.e. low losses. In Eq.~(\ref{maxwell}), $\sigma^{''}_{xx}$ ($\sigma^{''}_{yy}$) is the imaginary part of the $xx$ ($yy$) component of the conductivity tensor.
Substituting the Drude conductivity Eq.~(\ref{drude}) into Eq.~(\ref{maxwell}) and noting that the second term in r.h.s. can be neglected in most practical situations, we arrive at the classical plasmon dispersion of anisotropic 2DEG along the $i=x,y$ direction, $\omega^{i}_\mathrm{p}({\bf q})=\sqrt{(2\pi ne^2/m_i)|{\bf q}|}$. In the limit ${\bf q}\rightarrow 0$, the dependence $\omega_\mathrm{p} \sim \sqrt{|{\bf q}|}$ is universal for 2D, and does not depend on the energy dispersion of electrons. This allows for the existence of low-energy plasmons in 2D.

The same expression for the plasmon dispersion of 2DEG can be obtained within RPA, using Eq.~(\ref{rpa}). In this case, the energies of plasmon excitations are calculated from zeros of the dielectric function $\varepsilon(\omega,{\bf q})$, whereas the region of Landau damping is determined by the condition $\mathrm{Im}[\Pi(\omega,{\bf q})]\neq 0$.
Approximating the polarization function in the long-wavelength limit as \cite{Rodin2015}
\begin{equation}
\Pi(\omega,{\bf q}\rightarrow 0) \simeq \frac{g(\mu)\mu}{\omega^2}\left(\frac{q_x^2}{m_x} + \frac{q_y^2}{m_y}\right)    ,
\label{pi-longw}
\end{equation}
and using the bare 2D Coulomb interaction $v({\bf q})=2\pi e^2/|{\bf q}|$, we immediately obtain the following plasmon dispersion of anisotropic 2DEG from the solution $\mathrm{Re}[\varepsilon(\omega_\mathrm{p},{\bf q})]=0$,
\begin{equation}
    \omega_{\mathrm{p}}({\bf q}) \simeq \sqrt{\frac{g_s e^2\mu |{\bf q}|}{\hbar^2}} \left[ \sqrt{\frac{m_x}{m_y}} \mathrm{cos}^2\phi + \sqrt{\frac{m_y}{m_x}}\mathrm{sin}^2\phi \right]^{1/2},
    \label{wpl-2deg}
\end{equation}
where $g_s=2$ is the spin degeneracy factor, and $\phi$ is the angle of plasmon propagation relative to the $x$ axis. Interestingly, the region where the plasmon is undamped by the electron-hole continuum, i.e. where the imaginary part of the polarization function vanishes, depends on the direction. Assuming $m_x>m_y$, in the frequency region
\begin{equation}
 \frac{\hbar^2|{\bf q}|^2}{2m_x} + 2\sqrt{\mu}\frac{\hbar |{\bf q}|}{\sqrt{2m_x}}  <  \hbar \omega < \frac{\hbar^2|{\bf q}|^2}{2m_y} + 2\sqrt{\mu}\frac{\hbar |{\bf q}|}{\sqrt{2m_y}}
\end{equation}
the plasmon would be damped along the heavy mass ($x$) direction and undamped along the light mass ($y$) direction.

Figure~\ref{epsilon_eels}(b) shows the electron energy loss function \cite{vignale_book,vk_book} $L({\bf q},\omega)=-\mathrm{Im}[1/\varepsilon(\omega,{\bf q})]$ calculated beyond the long-wavelength limit by Prishchenko \emph{et al.} \cite{Prishchenko2017} for a moderately-doped single- and bilayer BP using a four-site TB model in combination with full Coulomb interaction. In both single- and bilayer cases one can see a prominent dispersion $\omega_\mathrm{p} \sim \sqrt{|{\bf q}|}$ at small ${\bf q}$, yet with a different scaling along the $x$ and $y$ directions. Apart from the standard square-root branch, one can see a damped acoustic branch dispersing as $\omega_\mathrm{p} \sim |{\bf q}|$. This mode originates from the finite thickness of BP, making the Coulomb potential dependent on the vertical distance $z$, $v({\bf q}) \sim e^{-|{\bf q}|z}$, resulting in a coupling between the (sub)layers and the emergence of collective modes \cite{Rodin2015}. In the context of BP, the existence of these modes were numerically demonstrated for bilayer BP in Refs.~\onlinecite{Jin2015,SaberiPouya2016}. It is also worth noting that for bilayer BP, the loss function in Fig.~\ref{epsilon_eels}(b) exhibits a weakly dispersed optical mode around $0.6$ eV, which originates from the interband transitions, and corresponds to out-of-phase oscillations of the charge density between the layers. The problem of anisotropic plasmons in monolayer BP was also addressed fully from first principles by Ghosh \emph{et al.} \cite{Ghosh2017}, with the full dielectric matrix calculated in the basis of plane waves. The advantage of such consideration is a more accurate description of the screening effects outside of the small-${\bf q}$ region, as well as the ability of capture local-field effects, typically ignored in the calculations based on TB or continuous models. Such effects, however, do not lead to a considerable modification of the plasmon spectrum of BP.

\begin{figure}[t]
\centering
\mbox{
\includegraphics[width=1.0\linewidth]{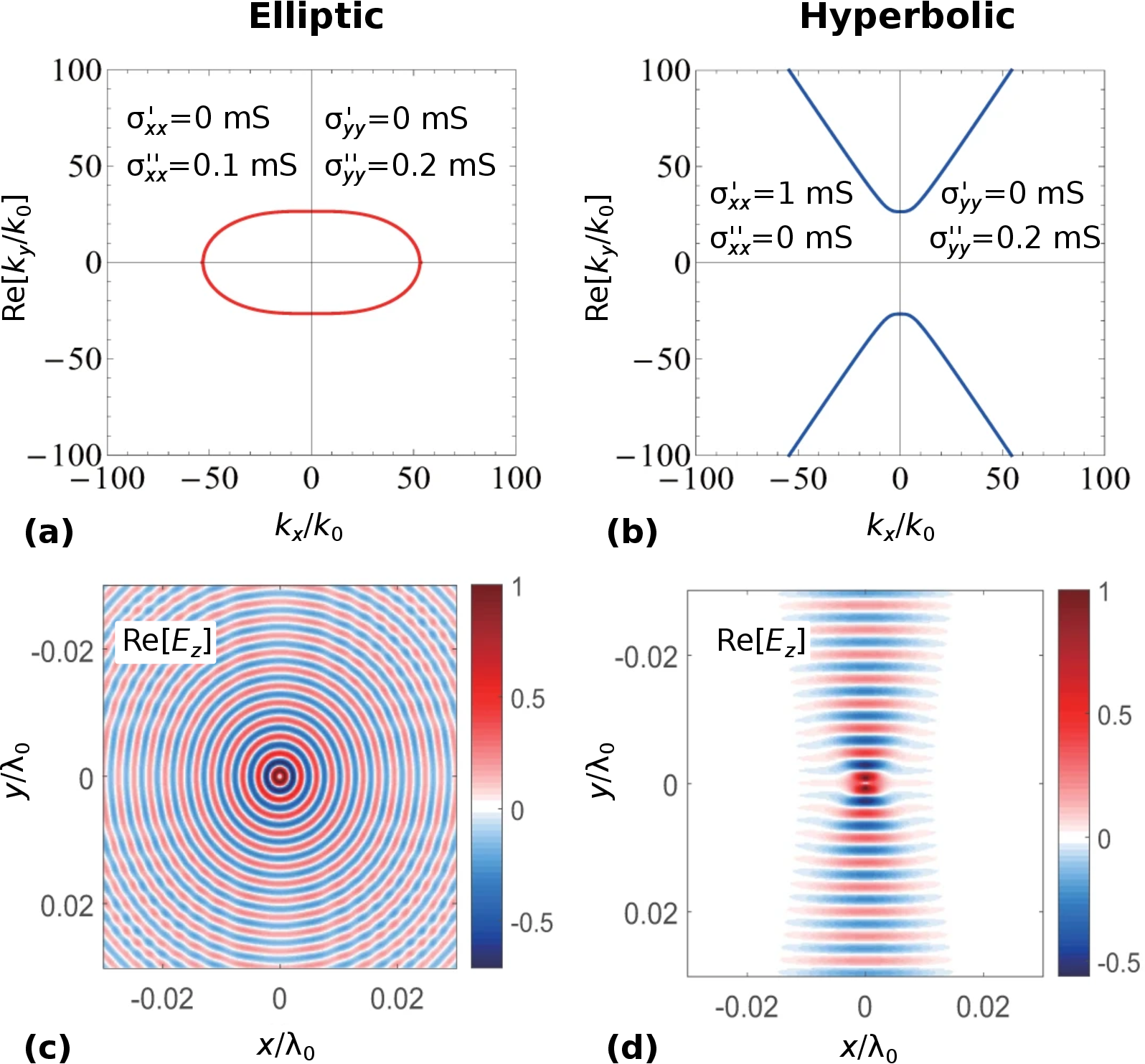}
}
\caption{
Illustration of the elliptic and hyperbolic regimes of plasmon propagation in anisotropic 2D materials. (a) and (b) show examples of elliptic and hyperbolic dispersions of plasmons obtained with the conductivity values given in the inset. (c) and (d) show the corresponding ((c) to (a) and (d) to (b)) spatial distributions of the electric field $\mathrm{Re}[E_z]$ induced by a $z$-oriented dipole. $k_0$ and $\lambda_0$ are the free-space wave number and wavelength, respectively. Adapted figure with permission from Ref.~\onlinecite{Chang2022}.
}
\label{hyper-plasmons}
\end{figure}

Anisotropic 2D materials provide new opportunities to achieve optical functionalities absent in their isotropic counterparts. The prominent example of a qualitatively new effect related to anisotropic 2D materials is their ability to host hyperbolic plasmons. From the dispersion relation (\ref{maxwell}), one can see that the isoenergy surface of plasmons $\omega(q_x,q_y)=\mathrm{const}$ is given by a closed ellipse provided that $\sigma_{xx}^{''}\sigma_{yy}^{''}>0$. In the opposite situation $\sigma_{xx}^{''}\sigma_{yy}^{''}<0$, the isoenergy surface is a hyperbola, allowing for the so-called hyperbolic regime of plasmon propagation \cite{Nemilentsau2016}. In this regime, the group velocity is predominantly focused along the directions normal to the hyperbola asymptotes, given by the angle $\mathrm{tan}\,\theta = \pm \sqrt{-\sigma_{xx}^{''}(\omega)/\sigma_{yy}^{''}(\omega)}$. The isoenergy surface is now open, meaning that large wave vectors can be supported, indicating the possibility of achieving highly localized plasmon beams.
For $\sigma^{''}_{yy}\gg\sigma^{''}_{xx}$ the plasmon propagation is strongly favored along the $y$ direction, which usually occurs when $\sigma^{''}_{xx} \approx 0$. Fig.~\ref{hyper-plasmons} shows an example of the plasmon energy dispersion in the elliptic and the hyperbolic regime with $\sigma^{''}_{xx} \approx 0$, and the 
corresponding field distributions launched by a dipole polarized in the $z$ direction. In the hyperbolic regime one can see the formation of a highly confined wavefront, in stark contrast to the elliptic regime.
The direction of plasmon propagation in the hyperbolic regime can be strongly controlled by frequency as well as by the Fermi energy. 
In practice, however, nonlocality effects stemming from the ${\bf q}$-dependence of the optical conductivity are expected to limit the confinement of plasmons in anisotropic materials \cite{CorreasSerrano2016}. Nevertheless, hyperbolic plasmonics of intrinsically anisotropic 2D materials such as BP opens up new perspectives for nanophotonic applications \cite{Nemilentsau2016,CorreasSerrano2017,vanVeen2019,Chang2022,Jia2022}. From the theoretical side, there have been a number of predictions for other materials that could potentially host hyperbolic plasmons. Among them are $\beta$-CP \cite{Dehdast2021}, CuB$_3$ and CuB$_6$ \cite{Geng2022}, NaW$_2$O$_2$Br$_6$ \cite{Enhui2022}, and ReS$_2$ \cite{Kiran2023}.
Experimentally, natural hyperbolic surface plasmons have been observed on thin films of WTe$_2$ in the micrometer frequency range \cite{Wang2020}.

Up to now, we have considered the polarizability of anisotropic 2DEG with a parabolic dispersion. The polarizability of anisotropic Dirac fermions was discussed in the context of strained graphene \cite{Pellegrino2011}. A more general analysis of density-density response in anisotropic Dirac systems was done by Hayn \emph{et al.} \cite{Hayn2021}. It turns out that the irreducible polarizability can be written in this case as 
\begin{equation}
    \Pi(\omega,{\bf q}) = \Pi^{\mathrm{iso}}(\omega,{\bf Q}),
\end{equation}
where $\Pi^{\mathrm{iso}}(\omega,{\bf q})$ is the well-known polarizability of isotropic Dirac fermions \cite{Katsnelson-Book} and ${\bf Q}$ is the wave vector with magnitude rescaled by a direction-dependent factor as
\begin{equation}
    |{\bf Q}|=|{\bf q}|\sqrt{\frac{v_y}{v_x}\mathrm{cos}^2\,\phi + \frac{v_x}{v_y}\mathrm{sin}^2\,\phi}.
\end{equation}
In the long-wavelength limit and at $\omega \gg |{\bf q}|\sqrt{v_xv_y}$, the polarizability then reads \cite{Hayn2021}
\begin{equation}
    \Pi(\omega,{\bf q}\rightarrow 0) \simeq \frac{g(\mu)}{2(\hbar \omega)^2} \left( v_y^2q_x^2 + v_x^2q_y^2 \right),
\label{piD-longw}
\end{equation}
which has the structure similar to Eq.~(\ref{pi-longw}). From Eq.~(\ref{piD-longw}), one immediately recovers the plasmon dispersion of anisotropic Dirac system,
\begin{equation}
    \omega_{\mathrm{p}}({\bf q}) \simeq \sqrt{\frac{g_se^2\mu |{\bf q}|}{\hbar^2}}\left[ \frac{v_y}{v_x}\mathrm{cos}^2\phi + \frac{v_x}{v_y}\mathrm{sin}^2\phi \right]^{1/2}.
    \label{wpl-dirac}
\end{equation}

The problem of dielectric screening and plasmons in anisotropic Dirac materials with tilted Dirac cones was addressed in Refs.~\onlinecite{Nishine2010,Sadhukhan2017,Jafari2018,Mojarro2022}. The tilt modifies the plasmon dispersion in such a way that the contributions from the two valleys become different. For each valley, the isoenergy surface is now asymmetric along the tilt axis, which might result in unusual effects in case of valley polarization. Besides, the tilt leads to an enhancement of the static screening as well as to a nonmonotonic behavior of the dynamical polarization, giving rise to cusps in the plasmon dispersion at the boundary of the electron-hole continuum \cite{Nishine2010,Jafari2018}. Overall, the tilt induces an additional source of anisotropy and may considerably affect optical response and screening properties of Dirac materials. Particularly, it can extend the frequency region that permits the existence of hyperbolic plasmons \cite{Mojarro2022}. Possible applications and experimental studies on plasmons in BP and other 2D anisotropic materials are reviewed in Ref.~\onlinecite{Wang2019}.

Dynamical polarizability and plasmons in 2D systems with merging Dirac points was analyzed in detail by Pyatkovskiy and Chakraborty \cite{Pyatkovskiy2016}. They considered a general Hamiltonian, Eq.~(\ref{H_expand}) under assumption $v_y^2/\Delta \gg m_y^{-1}$, and obtained analytical expressions for the polarizability. In the limit $q_y/\sqrt{2m_y}, v_xq_x \ll \omega \ll \mu$ a simplified expression for the polarizability can be expressed as
\begin{equation}
    \Pi({\omega,{\bf q}}) \simeq \frac{g_s\sqrt{\mu}\sqrt{2m_y}}{4\pi^2v_x(\hbar\omega)^2}\left[ v_x^2q_x^2 f_x\left(\frac{\Delta}{\mu}\right) + \frac{\mu}{2m_y}q_y^2f_y\left(\frac{\Delta}{\mu}\right) \right],
    \label{pi-general}
\end{equation}
yielding the plasmon dispersion
\begin{eqnarray}
\notag
        \omega_{\mathrm{p}}({\bf q}) \simeq \sqrt{\frac{g_se^2\mu |{\bf q}|}{2\pi\hbar^2}}\left[ \frac{\sqrt{\mu}}{v_x\sqrt{2m_y}} f_y\left(\frac{\Delta}
        {\mu}\right)\mathrm{cos}^2\phi \right. \\ 
        \left. + \frac{v_x\sqrt{2m_y}}{\sqrt{\mu}}f_x \left( \frac{\Delta}{\mu} \right) \mathrm{sin}^2\phi \right] ^{1/2}.
        \label{wpl-general}
\end{eqnarray}
Here, $f_x(\Delta/\mu)$ and $f_y(\Delta/\mu)$ are analytical functions that can be expressed in terms of the complete elliptic integrals \cite{Pyatkovskiy2016}. For well-separated Dirac cones, i.e. when $-\Delta \gg \mu$, one has $f_x \simeq \pi \sqrt{|\mu/\Delta|}$ and $f_y \simeq 4\pi\sqrt{|\Delta/\mu|}$, and Eq.~(\ref{pi-general}) reduces to the polarizability of anisotropic Dirac system in the long-wavelength limit, Eq.~(\ref{piD-longw}). In turn, the plasmon dispersion, Eq.~(\ref{wpl-general}) turns into Eq.~(\ref{wpl-2deg}). In the opposite limit $0 \ll \Delta \lesssim \mu$ (gapped semiconductor), the dispersion of electrons is parabolic at small ${\bf q}$ such that Eqs.~(\ref{pi-general}) and (\ref{wpl-general}) reduce to the polarizability and plasmon dispersion of anisotropic 2DEG, Eqs.~(\ref{pi-longw}) and (\ref{wpl-2deg}), respectively. The most unusual situation is realized upon merging of the Dirac points, i.e., at $\Delta=0$, corresponding to the case of a semi-Dirac system. This problem was addressed by Banerjee and Pickett \cite{Pickett2012}.
In this case, $f_x(0) \approx 5.8$ and $f_y(0) \approx 3.5$ are constants, and the plasmon dispersion becomes essentially different along the $x$ ($\phi=0$) and $y$ ($\phi = \pi/2$) directions, 
\begin{eqnarray}   
\notag
\omega_\mathrm{p}(q_x,0) \simeq \sqrt{\frac{g_se^2 |{\bf q}|}{2\pi\hbar^2}} \left[ f_x(0)v_x\sqrt{2m_y} \right]^{1/2}, \, \, \\
\omega_\mathrm{p}(0,q_y) \simeq \sqrt{\frac{g_se^2\mu^2 |{\bf q}|}{2\pi\hbar^2}} \left[ \frac{f_y(0)}{v_x\sqrt{2m_y}} \right]^{1/2}.
    \label{wpl-semidirac}
\end{eqnarray}
Remarkably, the plasmon frequency is independent of the chemical potential along the direction with linear dispersion of electrons ($y$), $\omega_{\mathrm{p}} \sim \mu^0$, and linearly dependent along the direction with quadratic dispersion $x$, $\omega_{\mathrm{p}} \sim \mu$. This behavior is closely related to the averaged Fermi velocities, which for a semi-Dirac system scale as  $\langle v_x^2 \rangle^{1/2} \propto v_x$ and $\langle v_y^2  \rangle^{1/2} \propto m_y^{-1/2}\sqrt{E}$ \cite{Pickett2012}. Unusual dependence of the plasmons frequency on the chemical potential in semi-Dirac materials suggests that doping could be used to control the degree of the anisotropy in these materials.

\subsection{\label{sec3c}Anisotropic excitons}

\begin{figure}[b]
\centering
\mbox{
\includegraphics[width=0.7\linewidth]{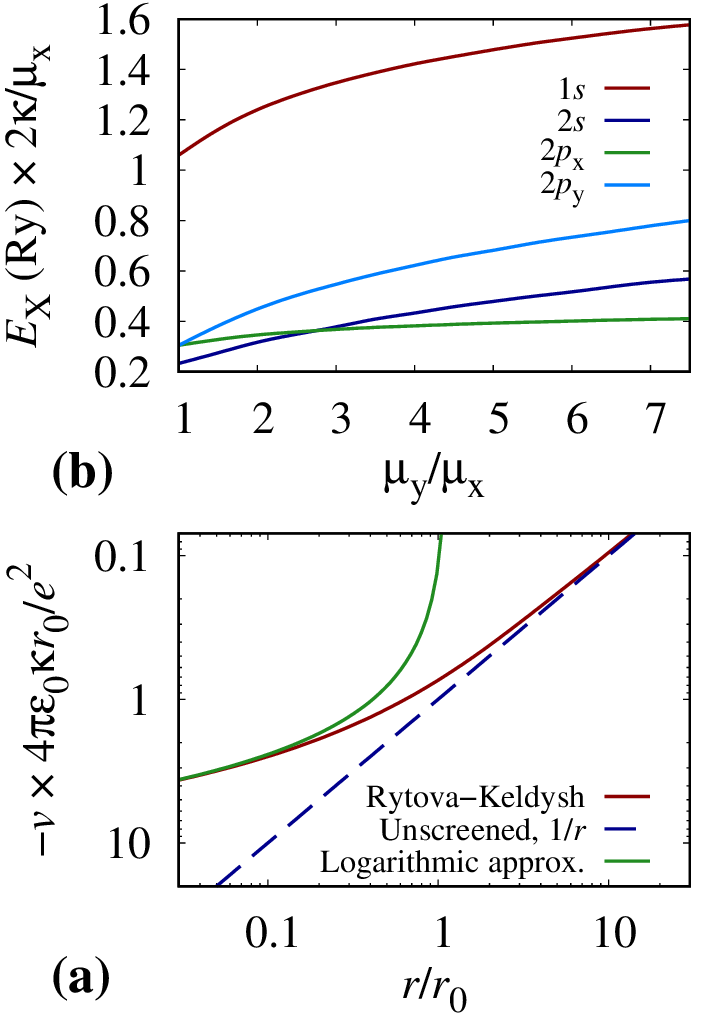}
}
\caption{(a) Electrostatic potential created by a point charge in 2D given by Eq.~(\ref{RK-potential}), and its approximations in the $r\gg r_0$ and $r\ll r_0$ limits given by Eqs.~(\ref{V_unscreen}) and (\ref{V_log}), respectively. (b) Exciton binding energies for the ground state and three excited states calculated within the effective mass approximation as a function of the reduced mass anisotropy $\mu_x/\mu_y$ (tabulated from Ref.~\onlinecite{Chaves2015_excitons}).
}
\label{RK-fig}
\end{figure}

The excitonic behavior of 2D anisotropic systems also exhibits a number of peculiarities compared to their isotropic analogs. The minimal model for the exciton dynamics in 2D is based on the effective mass approximation, which describes an anisotropic two-body system with an attractive potential. The corresponding Hamiltonian is given by
\begin{equation}
    H = \frac{p_x^2}{2\mu_x} + \frac{p_y^2}{2\mu_y} + v(|{\bf r}_{\mathrm e} - {\bf r}_\mathrm{h}|),
    \label{exc_hamilt}
\end{equation}
where ${\bf p}=(p_x,p_y)$ is the momentum operator related to the center of mass of particles located at positions ${\bf r}_{\mathrm e}$ and ${\bf r}_\mathrm{h}$. Since the mass of each particle is anisotropic, there are two different reduced masses in the $x$ and $y$ directions,
\begin{equation}
    \mu_{x(y)} = \frac{m^\mathrm{e}_{x(y)}m^\mathrm{h}_{x(y)}}{m^\mathrm{e}_{x(y)}+ m^\mathrm{h}_{x(y)}}
\end{equation}
with $m^\mathrm{e}_{x(y)}$ and $m^\mathrm{h}_{x(y)}$ being the effective masses of electrons and hole in the $x$ ($y$) direction. The first two terms in the Hamiltonian Eq.~(\ref{exc_hamilt}) correspond to the kinetic energy, while the last term $v(r)$ is the  attractive central potential created by a point charge in 2D, known as the Rytova-Keldysh potential \cite{Rytova,Keldysh}, 
\begin{equation}
v(r) = -\frac{e^2}{8\varepsilon_0 \kappa r_0}\left[ H_0\left( \frac{r}{r_0} \right) - Y_0\left(  \frac{r}{r_0}  \right) \right] ,
\label{RK-potential}
\end{equation}
where $\varepsilon_0$ is the vacuum permittivity, $H_0$ is the zeroth-order Struve function, $Y_0$ is the zeroth-order Bessel function of the second kind, $\kappa=(\varepsilon_1 + \varepsilon_2)/2$ is the average dielectric constant of the media surrounding the 2D material from both sides. $r_0=d\varepsilon/2\kappa$ plays the role of the screening length, and it is related to the effective layer thickness $d$ and its 
effective dielectric constant $\varepsilon$. Importantly, the latter is supposed to be isotropic, therefore, strictly speaking, its use for anisotropic materials is not fully justified. Nevertheless, the expression (\ref{RK-potential}) is frequently used. 

The length $r_0$ defines a crossover between the conventional Coulomb potential at long distances \cite{Keldysh},
\begin{equation}
    v(r\gg r_0) \simeq -\frac{e^2}{4\pi\varepsilon_0\kappa}\frac{1}{r}
    \label{V_unscreen}
\end{equation}
and its logarithmic divergence at short distances,
\begin{equation}
    v(r\ll r_0) \simeq \frac{e^2}{4\pi\varepsilon_0\kappa}\frac{1}{r_0}\left[ \mathrm{ln}\left(\frac{r}{2r_0} \right) + \gamma\right],
    \label{V_log}
\end{equation}
where $\gamma \approx 0.577$ is the Euler constant. From Eq.~(\ref{V_unscreen}) one can see that the electrons in 2D do not undergo a self-screening at large distances, which is a general result, independent of form of the energy dispersion \cite{Katsnelson-Book}. The form of the Rytova-Keldysh potential as well as its limiting approximations are shown in Fig.~\ref{RK-fig}(a).
It is worth noting that the two-body problem of particles with non-parabolic dispersion (e.g., Dirac fermions) cannot be mapped into that of a single particle due to the absence of Galilean invariance \cite{Sabio2009}, which makes it difficult to construct an analytical theory for excitons in generic 2D materials. We also note the existence of phenomenological electron-hole interaction potentials different from Eq.~(\ref{RK-potential}) that effectively take into account material's thickness and many-body effects by fitting to the results of first-principles calculations \cite{Villegas2016}.

Even in the limit $r \gg r_0$, the anisotropic Hamiltonian (\ref{exc_hamilt}) cannot be diagonalized exactly contrary the isotropic case, which corresponds to the problem of 2D hydrogen atom \cite{Yang1991}. Rodin \emph{et al.} \cite{Rodin2014_excitons} proposed a change of variables that transforms the problem of anisotropic masses into the problem of an anisotropic potential with the degree of anisotropy characterized by the parameter $\beta=(\mu_y - \mu_x) / (\mu_y + \mu_x)$. After this reformulation, the Schr\"{o}dinger equation with the Hamiltonian (\ref{exc_hamilt}) can be efficiently solved using numerical methods \cite{Rodin2014_excitons}. Recently, Henriques and Peres \cite{Peres2020} developed a semi-analytical approach method based on the perturbative treatment of the potential energy with respect to the parameter $\beta$, offering perhaps the most computationally effective way to solve Eq.~(\ref{exc_hamilt}) in systems with moderate anisotropy. An approximate solution to the problem can be obtained by variational methods. 
A variational solution for the ground-state exciton wave function can be chosen in analogy to the isotropic problem in the form
\begin{equation}
\phi(x,y) = \sqrt{\frac{2}{\pi}}(a_xa_y)^{-1/2}\mathrm{exp}\left[ -\sqrt{\left(x/a_x\right)^2 + \left(y/a_y\right)^2} \right],
\label{phi_var}
\end{equation}
where $a_x$ ($a_y$) characterizes spatial extension of
the exciton along the $x$ ($y$) direction. With this variational wave function and the Hamiltonian (\ref{exc_hamilt}) one can evaluate the expectation value of the ground-state energy $E_0=\langle \phi |H| \phi \rangle$, which approximates the exciton binding energy $E_\mathrm{X}\simeq E_0$. 
Upon minimization of this energy with respect to the parameters $a_x$ and $a_y$, the authors of Ref.~\onlinecite{CastellanosGomez2014} obtained numerically the optimal set of parameters for monolayer BP. A more general consideration and benchmarking of the variational solution (\ref{phi_var}) was carried out by Prada \emph{et al.} \cite{Prada2015}. In the limits $r\gg r_0$ and $r \ll r_0$ they found analytical expressions for $a_x$, $a_y$, and $E_\mathrm{X}$. In what follows, we present the results only for the regime $r\ll r_0$ as this case reflects the situation typical to 
2D semiconductors like BP and MoS$_2$, where the screening length is of the order of few nm, whereas the exciton radii are considerably smaller. In this regime, the minimal spatial extensions as functions of the anisotropy parameter $\lambda=a_y/a_x$ read
\begin{equation}
    a_x(\lambda) = \sqrt{a_0m_0 \frac{\kappa r_0}{\mu_{xy}(\lambda)}},\text{ } a_y(\lambda) = \sqrt{a_0m_0 \frac{\kappa r_0 \lambda^2}{\mu_{xy}(\lambda)}},
\end{equation}
where $2\mu^{-1}_{xy}(\lambda) = \mu_x^{-1} + \lambda^{-2}\mu_y^{-1}$, and $a_0$ and $m_0$ correspond to the Bohr radius and free-electron mass, respectively. The corresponding exciton energy reads
\begin{equation}
E_\mathrm{X}(\lambda) = \frac{e^2}{4\pi \varepsilon_0 \kappa}\frac{1}{r_0}\left[ \frac{3}{2} + \mathrm{ln}\left( \frac{a_x(\lambda)}{4r_0}\frac{\lambda + 1}{2}\right) \right].
\end{equation}
In turn, the variational anisotropy of the exciton extension is given by
\begin{equation}
\lambda = \frac{a_y}{a_x} = \left( \frac{\mu_x}{\mu_y} \right)^{1/3}.
\label{lambda-exc}
\end{equation}
For $r_0=20a_0$, the deviation between the analytical result given above and the exact numerical evaluation of $E_\mathrm{X}$ is only around 10\% \cite{Prada2015}. Recently, Gomes \emph{et al.} \cite{Gomes2022} extended the variational method to the case of multilayer 2D semiconductors with finite thickness by evaluating corrections to the potential Eq.~(\ref{RK-potential}), as well as provided closed-form expressions for the energy of the first excited states.

From Eq.~(\ref{lambda-exc}) one can see that spatial extension of the exciton along a given direction is inversely proportional to the effective mass in this direction. In other words, the exciton wave function becomes more extended along the direction of the smaller mass. Another effect of the band anisotropy is related to lifting the degeneracy of the states with nonzero angular momentum, e.g., $2p_x$ and $2p_y$. Unlike the standard Coulomb problem with the unscreened potential $v(r)\sim 1/r$, the logarithmic part of the Rytova-Keldysh potential ensures that the degeneracy of the 2$s$ and 2$p$ states is lifted even in the isotropic case. For anisotropic systems, all three excited states turn out to have different energies \cite{Rodin2014_excitons,Gomes2022,Chaves2015_excitons}. The dependence of the 1$s$, 2$s$, 2$p_x$, and 2$p_y$ excitonic energies on the reduced mass anisotropy is shown in Fig.~\ref{RK-fig}(b). Apart from increasing the splitting between the excited states, one can see that the anisotropy increases absolute values of the exciton binding energies.

In the context of BP, San-Jose \emph{et al.} \cite{San-Jose2016_funnel} found another remarkable anisotropy-related effect associated with lattice deformations. Particularly, in the presence of inhomogeneous strain excitons in BP were predicted to accumulate away from the regions of high tensile strain, accompanied by highly anisotropic exciton flow. This effect was dubbed the inverse funnel effect, in contrast to the normal funnel effect typical to isotropic 2D materials such as MoS$_2$, where excitons tend to accumulate isotropically around the regions of high tensile strain. The inverse funnel effect is not generic to anisotropic 2D materials, but related to peculiarities of the BP crystal structure, leading to the unusual sign of gap modulation with strain ($\partial \Delta / \partial u_{ii}>0$). The role of uniform strain in the excitonic properties of anisotropic 2D materials were considered in Ref.~\onlinecite{Seixas2015} within the effective mass approximation.

A more general, yet computationally demanding approach to study excitons in 2D systems with arbitrary energy dispersion is to use \emph{ab initio} calculations based on the solution of the many-body Bethe–Salpeter equation (BSE), which is usually combined with first-principles calculations within the $GW$ approximation \cite{Rohlfing2000}. There have been a large number of studies applying this method in its different variations to monolayer and multilayer BP, as well as to other anisotropic 2D materials \cite{Tran2014,Tran2015,Choi2015,Zhong2015,Qiu2017,Ferreira2017,Gerber2018,Junior2019_kp,Arra2019}. The problem of excitons in BP was also addressed using diffusive quantum Monte Carlo methods \cite{Hunt2018}, which might be also applicable to estimate the binding energies of biexcitons. Finally, we mention the work of Deilmann and Thygesen \cite{Deilmann2018}, who studied positive and negative charged trions in BP using an extension of the $GW$-BSE methodology.

Experimentally, highly anisotropic excitons were observed in monolayer and few-layer BP by polarization-resolved photoluminescence and infrared absorption spectroscopy \cite{Wang2015,Surrente2016,Zhang2018}. Using the reflection and photoluminescence measurements, Tian \emph{et al.} \cite{Tian2020} observed the excitonic series in multilayer BP with a pronounced deviation from the hydrogenic series. Carr\'{e} \emph{et al.} \cite{Carre2024} performed a systematic analysis of the photoluminescence spectra of numerous mechanically-exfoliated BP flakes spanning a wide range of thicknesses, and found that the emission process is dominated by the defects introduced by the exfoliation process. Moreover, for samples with intermediate thickness they show that the evolution of the excitonic energy versus thickness follows an inverse square law, originating from the confinement effects, being mostly independent of the dielectric environment.

Besides BP, anisotropic excitons were also observed in atomically thin ReS$_2$ and ReSe$_2$ \cite{Sim2016,Arora2017,Sim2018,JWang2020}.
Recently, excitons were studied, both experimentally and theoretically, in highly anisotropic system, van der Waals magnet CrSBr \cite{Ruta2023}. 
In this compound, $\sigma''(\omega)$ turns out to be negative along one of the in-plane directions at the photon energies around $\hbar \omega = 1.40$ eV, which results in the formation of hyperbolic electromagnetic waves. Since excitonic contribution to optical conductivity is dominant in this energy range, these electromagnetic waves were referred to as hyperbolic excitons. Their spectral weight turns out to be very sensitive to magnetic order, dramatically increasing at low temperatures upon formation of antiferromagnetic order between the layers. 

\section{\label{sec4}Charge carrier scattering and electronic transport}

\subsection{\label{theory_scatter}Theory of anisotropic scattering}

In the semiclassical theory, electron transport can be described by the mechanism of quasiparticle propagation \cite{Allen2006}. The current density ${\bf j}$ occurring in response to the electric field is determined by the non-equilibrium distribution function $f({\bf k})$
\begin{equation}
    {\bf j} = -e\sum_{\bf k}{\bf v}({\bf k}) [f({\bf k}) - f_0({\bf k})],
    \label{current}
\end{equation}
where ${\bf v}({\bf k})=\hbar^{-1}\partial \varepsilon_{\bf k} / \partial {\bf k}$ is the group velocity and $f_0({\bf k})$ is the equilibrium (Fermi-Dirac) distribution function. In the Boltzmann transport approach, the distribution function $f({\bf k})$ is determined by the well-known Boltzmann transport equation \cite{Allen2006,Ziman-Book}. Assuming a homogeneous dc electric field in the absence of magnetic field, the current flow can be described by a stationary distribution, which satisfies the equation
\begin{equation}
    -\frac{e}{\hbar}{\bf E}\cdot \nabla_{\bf k}f({\bf k}) =
    I(f)
    \label{boltz1}
\end{equation}
where $I(f)$ is the collision integral expressed as
\begin{equation}
I(f)=2\pi \sum_{{\bf k}'}\delta(\varepsilon_{\bf k}-\varepsilon_{{\bf k}'})\langle |V_{{\bf k}{\bf k}'}|^2\rangle [f({\bf k})-f({{\bf k}'})],
\label{boltz2}
\end{equation}
where $V_{{\bf k}{\bf k}'}=\langle \psi_{\bf k}|V|\psi_{{\bf k}'} \rangle$ is the scattering amplitude, and $\langle ... \rangle$ denotes statistical average. Here we consider only the case of elastic scattering by static potentials because all main effects of anisotropy can be clearly seen already from this simplest, but still important, situation. The distribution $f({\bf k})$ can be expanded around the equilibrium following the \emph{ansatz} \cite{Allen2006}
\begin{equation}
    f({\bf k}) = f_0({\bf k}) +e\frac{\partial f_0}{\partial \varepsilon} [{\bf v}({\bf k})\cdot {\bf E}] \, \tau({\bf E},{\bf k}),
    \label{boltz3}
\end{equation}
which is equivalent to a field-dependent shift of energies in the equilibrium distribution $f(\varepsilon_{\bf k}) \rightarrow f_0(\varepsilon_{\bf k} + e[{\bf v}({\bf k})\cdot {\bf E}] \, \tau({\bf E},{\bf k}) )$. Here, $\tau({\bf E},{\bf k})$ is the transport relaxation time. While this shift is independent of the field direction for an isotropic energy dispersion, this is not the case for an anisotropic one. As a result, the relaxation time is dependent on the direction of applied electric field. Assuming the electric field is applied along the $x$ direction, we arrive at the following implicit equation for the relaxation time after combining Eqs.~(\ref{boltz1}), (\ref{boltz2}), and (\ref{boltz3}), 
\begin{equation}
    v_x({\bf k}) = 2\pi \sum_{{\bf k}'}\delta(\varepsilon_{\bf k} - \varepsilon_{{\bf k}'}) \langle |V_{{\bf k}{\bf k}'}|^2\rangle \left[v_x({\bf k})\tau_x({\bf k}) - v_x({\bf k}')\tau_x({\bf k}')\right].
    \label{tau_closed}
\end{equation}
For isotropic systems, $\tau$ is independent on the direction of the vector ${\bf k}$ and from Eq.~(\ref{tau_closed}) we immediately obtain the well-known expression for the inverse relaxation time
\begin{equation}
    \tau^{-1}_{\mathrm{iso}} = \frac{2\pi}{\hbar}\sum_{{\bf k}'}\delta(\varepsilon_{\bf k} - \varepsilon_{{\bf k}'}) \langle |V_{{\bf k}{\bf k}'}|^2\rangle (1-\mathrm{cos}\theta_{{\bf k}{\bf k}'}),
\end{equation}
where $\theta_{{\bf k}{\bf k}'}$ is the scattering angle between vectors ${\bf k}$ and ${\bf k}'$.

The formula for the $x$ component of the conductivity can be obtained from Eqs.~(\ref{current}) and (\ref{boltz3}) \cite{Allen2006}
\begin{equation}
    \sigma_{xx} = e^2\sum_{\bf k}v^2_x({\bf k})\tau_{x}({\bf k})\left( -\frac{\partial f_0}{\partial \varepsilon_{\bf k}} \right).
\end{equation}
For moderately anisotropic systems, one can use the standard variational solution which yields conductivity close to what follows from the exact solution of Boltzmann equation. In the variational approach \cite{Ziman-Book}, the conductivity reads $\sigma_{xx} = e^2 
g(E_\mathrm{F}) \langle v_x^2 \rangle \tau_{xx}$, where $\tau_{xx}$ is the Fermi-averaged transport relaxation time
calculated as
\begin{equation}
\tau_{xx}^{-1} = \frac{2\pi}{\hbar}\frac{\sum_{{\bf k}{\bf k}'}\delta(\varepsilon_{\bf k} - \varepsilon_{{\bf k}'})\left(-\frac{\partial f_0}{\partial \varepsilon_{\bf k}}\right)(v^{{\bf k}}_x-v_x^{{\bf k}'})^2\langle |V_{{\bf k}{\bf k}'}|^2 \rangle} {\sum_{\bf k}v^2_x({\bf k})\left(-\frac{\partial f_0}{\partial \varepsilon_{\bf k}}\right)}.
\label{tauxx-variational}
\end{equation}
In Sec.~\ref{sec_chg-imp} below, we will consider an example for which we compare the scattering rate calculated within the variational approach with the result of numerical solution of the Boltzmann equation.
It is worth noting that the solution of the Boltzmann equations within the variational approach is equivalent to the Kubo-Nakano-Mori formula for the conductivity, which is based on the Kubo approach yet without symmetry constraints \cite{Katsnelson-Book}.

Let us now consider specific examples of typical scattering potentials.

\subsection{\label{sec4b}Scattering by random disorder}
A pointlike uncorrelated disorder with the random potential $V({\bf r})$ is characterized by the conditions
\begin{equation}
    \langle V({\bf r}) \rangle=0 \text{  and  } \langle V({\bf r})V({\bf r}') \rangle=\gamma\delta({\bf r} - {\bf r}' ),
\end{equation}
where $\gamma$ determines the disorder strength. Physically, it corresponds to the case of short-range scattering potential. For the case of Dirac systems, the contribution of long-range scatterers, such as charge impurities, ripples, or resonant scattering centers is much more important \cite{Katsnelson-Book} but even for this situation it may be instructive to start with the simplest situation of point-like scatterers, to see the difference between isotropic and anisotropic case. Long-range (that is, Coulomb) potential will be considered in the next subsection. For a single quadratic band the scattering matrix element is independent of the scattering wave vector ${{\bf q}={\bf k}-{\bf k}'}$, and is given by $\langle | V_{{\bf k}{\bf k}'} |^2\rangle=\gamma$. In this situation, the transport scattering time in Eq.~(\ref{tau_closed}) is independent of ${\bf k}$ and coincides with the electronic scattering time $\tau = \tau^\mathrm{e}$ defined in terms of the Fermi golden rule,
\begin{equation}
\tau^{\mathrm{e}}(E) = \frac{\hbar}{\pi \gamma g(E)},
\label{tau_bare}
\end{equation}
Therefore, for anisotropic systems with quadratic energy dispersion and random disorder, anisotropy of the transport properties is determined exclusively by the anisotropy of the Fermi surface. Specifically, the conductivity along the $x$ direction reads $\sigma_{xx} = (e^2\hbar/\pi \gamma)E_\mathrm{F}/2m_x$. This result is in agreement with numerical large-scale simulations of point defects within the TB model \cite{Yuan2015}.

The situation with anisotropic Dirac and semi-Dirac systems in the presence of random disorder is less trivial because of the angular dependence of the eigenstates overlap entering the scattering matrix. This problem was addressed in detail by Adroguer \emph{et al.} \cite{Montambaux2016}, where conductivity has been studied across the
topological transition between a semiconducting phase and a
Dirac phase. For an anisotropic Dirac system with well separated cones along the $y$ direction, in the limit $E \ll |\Delta|$ the electronic scattering time is also determined by the expression Eq.~(\ref{tau_bare}). However, the transport scattering time becomes anisotropic due to the intervalley scattering along the $y$ direction. 
Specifically, one obtains $\tau_x=2\tau^\mathrm{e}$ and $\tau_y=\tau^\mathrm{e}$ \cite{Montambaux2016}. This leads to the conductivities
\begin{equation}
    \sigma_{xx} = \frac{e\hbar^2}{\pi \gamma}v_x^2 \,\text{, }\, \sigma_{yy} = \frac{e\hbar^2}{2\pi \gamma}v_y^2,
\end{equation}
which are now not only determined by the anisotropy of the Fermi velocities, but also by the anisotropy of the transport relaxation time.
It is worth noting that $\sigma_{xx}$ is the same as the result obtained for the case of independent Dirac cones.

Close to the merging transition, i.e., when the Dirac cones are not well separated from each other, the electronic scattering time becomes anisotropic by acquiring angular dependence
\begin{equation}
    \tau^{\mathrm{e}}(E,\theta) = \frac{\hbar}{\pi \gamma g(E)} \left(1 + r\left(\frac{\Delta}{E}\right)\mathrm{cos}\,\theta\right)^{-1},
    \label{tau-e_ang}
\end{equation}
where $r(\delta)$ is a non-monotonic anisotropy function, whose explicit expression can be found in Ref.~\onlinecite{Montambaux2016}. In the limit considered above, $\delta=E/\Delta\ll -1$, $r(\delta)\rightarrow 0$ and Eq.~(\ref{tau-e_ang}) becomes Eq.~(\ref{tau_bare}). Beyond this limit, the transport time along $x$ undergoes a renormalization
$\tau_x=\lambda(\delta)\tau^{\mathrm{e}}$ with the factor $\lambda(\delta)\lesssim 3.6$, while the transport time along the $y$ direction remains unchanged, $\tau_y=\tau^{\mathrm{e}}$. At the transition point $\Delta=0$, $\lambda(0)\simeq 2.5$. More importantly, however, is a qualitative difference in the behavior of the averaged squared velocities, which leads to the following conductivity dependencies for a semi-Dirac system
\begin{equation}
\sigma_{xx}(E_\mathrm{F}) \simeq  1.5\, \frac{e^2\hbar}{\pi \gamma}v_x^2 \,\text{, }\,\sigma_{yy}(E_{\mathrm{F}}) \simeq 0.2\, \frac{e^2\hbar}{\pi \gamma}\frac{2E_{\mathrm{F}}}{m_y}.
\label{sigma-semidirac}
\end{equation}
A similar result has been obtained earlier in Ref.~\onlinecite{Banerjee2009} yet with a different numerical coefficients due to an approximate treatment of the scattering matrix.

The unusual conductivity behavior in semi-Dirac systems has been ascribed by Ziegler and Sinner \cite{Ziegler2017} to anomalous diffusion. Indeed, the semi-classical conductivity can be represented as proportional to the direction-dependent diffusion coefficient $D_{\alpha}$, i.e., $\sigma_{\alpha \alpha} = e^2 g(E_\mathrm{F})D_{\alpha}$ in accordance with the Einstein model. While for an isotropic energy dispersion $D_{\alpha}$ is energy-independent, from Eq.~(\ref{sigma-semidirac}) it follows that $D_{x}\sim E^{-1/2}$ and $D_{y}\sim E^{1/2}$. Such behavior has been interpreted in Ref.~\onlinecite{Ziegler2017} as a difference in the scaling of the long-time mean square displacement, $\langle ({\bf r}_i - {\bf r}_j)^2 \rangle \sim t^{\alpha}$. While for conventional diffusion $\alpha=1$, in the case of a semi-Dirac system one has $\alpha =3/2$ along the $x$ direction (linear dispersion) and $\alpha=1/2$ along the $y$-directions (quadratic dispersion), corresponding to the superdiffusive and subdiffusive behavior, respectively. It is worth noting that a similar unusual behavior has been predicted for thermal conductivity and thermoelectric
coefficients in semi-Dirac systems \cite{Mandal1,Mandal2}.

\begin{figure}[tbp]
\centering
\mbox{
\includegraphics[width=0.8\linewidth]{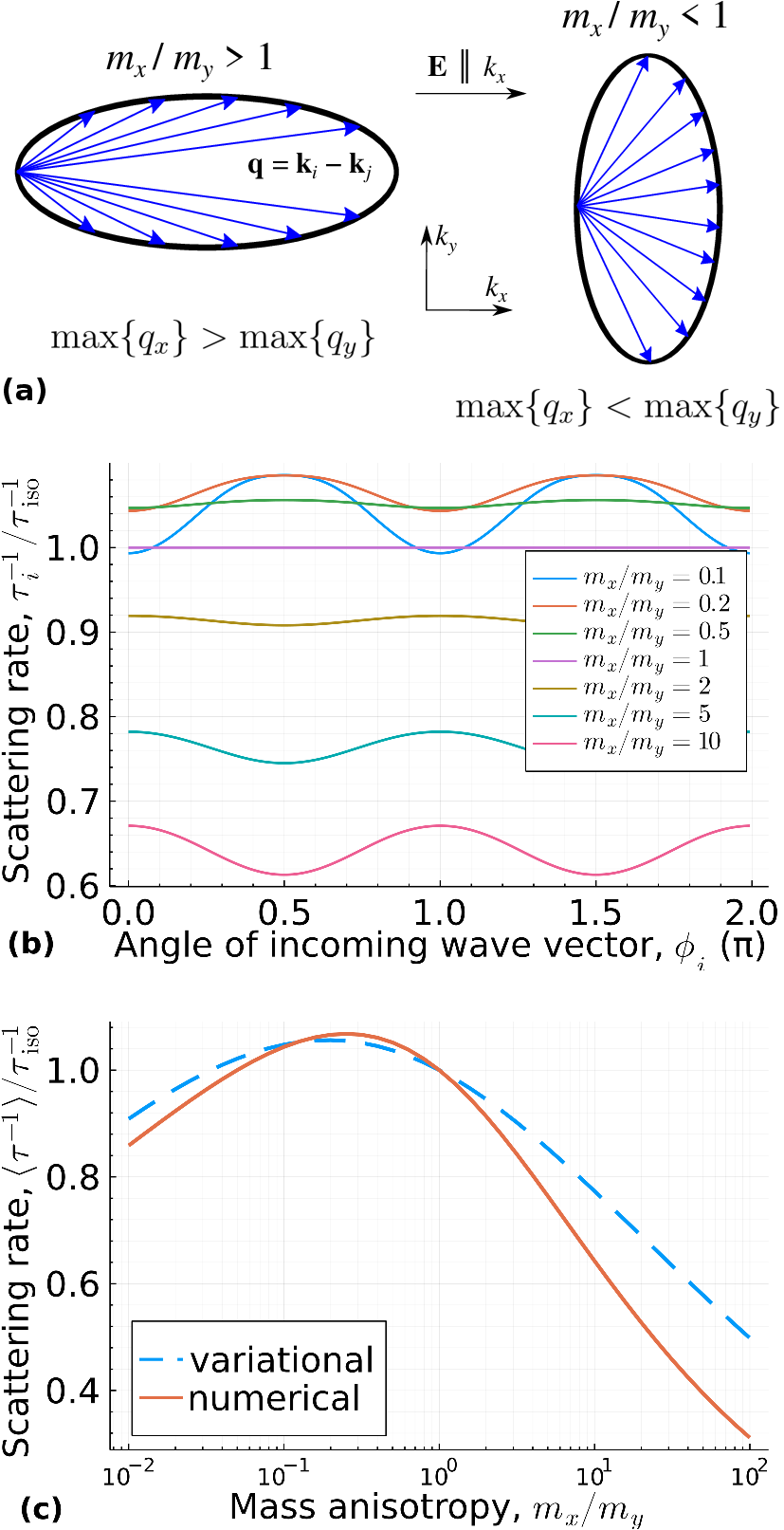}
}
\caption{(a) Schematic representation of the scattering process on an elliptical Fermi contour with two different effective mass ratios. (b) Scattering rate $\tau_x^{-1}({{\bf k}_i})$ calculated as a function of the incoming wave vector angle $\phi_i$ obtained from the numerical solution of the Boltzmann equation shown for different anisotropy ratios. (c) Scattering rate averaged over the Fermi contour $\langle \tau_x^{-1} \rangle$ calculated as a function of the effective mass anisotropy $m_x/m_y$ obtained from the numerical solution of the Boltzmann equation (solid line) and from its variational solution (dashed lines). In all cases DOS~$\sim \sqrt{m_xm_y}$ is kept constant for each anisotropy ratio considered. $\tau_{\mathrm{iso}}^{-1}$ is the scattering rate of an isotropic system with the same DOS. Charge impurity scattering potential in the form of Eq.~(\ref{charge-imp-pot}) is assumed in all cases with the Fermi energy $E_\mathrm{F}=0.1$ a.u.
}
\label{fig5_scattering}
\end{figure}

\subsection{\label{sec_chg-imp}Charge impurity scattering}

Let us now consider scattering on charged impurities. In this case, the scattering matrix element is explicitly dependent on the direction of the wave vector. Moreover, in the limit of long wave lengths, the scattering potential is isotropic, which is discussed in Sec.~\ref{plasmons}. The corresponding scattering matrix element can be written as \cite{RevModPhys.54.437}
\begin{equation}
    \langle |V_{{\bf q}}|^2 \rangle = 
    n_{\mathrm{imp}} |\langle \psi_{\bf k}| W |\psi_{{\bf k}'} \rangle|^2,
    \label{charge-imp-pot}
\end{equation}
where ${\bf q} = {\bf k} - {\bf k'}$,  $W({\bf q})=v({\bf q})e^{-q\cdot h}/\varepsilon({\bf q})$ is the screened Coulomb potential, $q = |{\bf q}|$ and $h$ is the vertical distance from the charged impurity to the 2D layer. As in Sec.~\ref{plasmons}, $v({\bf q})=2\pi e^2/q$ and $\varepsilon({\bf q})=1-v({\bf q})\Pi({\bf q})$ are the bare Coulomb potential and dielectric function, respectively. It is worth noting that $\langle |V_{{\bf q}}|^2 \rangle$ is dependent on the Fermi energy via the dependence of $\Pi({\bf q})$ [see Eq.~(\ref{polariz_2d})].
In the context of monolayer BP under the assumption of parabolic bands, the problem of charge impurity scattering was addressed by Liu \emph{et al.} in Ref.~\onlinecite{Ruden2016}.

Despite the fact that the scattering potential is isotropic, the scattering of electrons in an anisotropic band is strongly anisotropic. For simplicity, we restrict our consideration to the case of a parabolic band. Figure \ref{fig5_scattering}(a) shows schematic scattering on an elliptical Fermi surface for the cases $m_x/m_y>1$ and $m_x/m_y<1$. Assuming the electric field is applied along the $x$ direction, this direction corresponds to the preferable scattering direction. Indeed, from Eq.~(\ref{tauxx-variational}) one can see that for a parabolic band the scattering probability enters the integral with the factor $(v_x^{\bf k} - v_x^{{\bf k}'})^2\sim q_x^2$. Therefore, for $m_x/m_y > 1$ the scattering occurs on larger wave vectors ${\bf q}$ contrary to the case $m_x/m_y < 1$. Keeping in mind that $\langle |V_{\bf q}|^2 \rangle \sim q^{-2}$ for scattering on charged impurities, the overall scattering rate $\tau_{xx}^{-1}$ must be suppressed for $m_x/m_y>1$, i.e., when the Fermi contour is elongated along the field direction. 

Figure \ref{fig5_scattering}(b) shows the scattering rate $\tau_{xx}({\bf k}_i)$ calculated from the numerical solution of Eq.~(\ref{tau_closed}) as a function of $\phi_i$, which is the angle of incoming wave vector ${\bf k}_i=k_i\mathrm{cos}\,\phi_i$, for different $m_x/m_y$. Contrary to the isotropic case, $\tau^{-1}_{xx}({\bf k}_i)$ exhibits an oscillatory dependence on $\phi_i$. For $m_x/m_y>1$, $\tau^{-1}_{xx}$ has minima at $\phi_i=\pi/2$ and $\phi_i=3\pi/2$. However, for moderate anisotropies, such oscillations are not significant, being much smaller compared to the absolute scattering rate. 
  Fig.~\ref{fig5_scattering}(b) also clearly reflects the fact that the scattering rate is strongly dependent on the effective mass anisotropy. While $\tau^{-1}_{xx}$ decreases steadily for $m_x/m_y>1$, it demonstrates less trivial behavior for $m_x/m_y<1$. At $m_x/m_y<1$ averaged $\tau^{-1}_{xx}$ increases and exhibits a maximum at some critical $m_x/m_y$ determined by the Fermi energy and details of the scattering potential. Interestingly, at this point $\tau^{-1}_{xx}>\tau^{-1}_{\mathrm{iso}}$. As $m_x/m_y$ continues to decrease, $\tau^{-1}_{xx}$ decreases as well. This behavior is depicted more clearly in Fig.~\ref{fig5_scattering}(c), which displays evolution of the averaged scattering rate $\langle \tau^{-1}_{xx} \rangle$ with the anisotropy $m_x/m_y$. In Fig.~\ref{fig5_scattering}(c), we also show a comparison between $\tau_{xx}^{-1}$ calculated from numerical solution of the Boltzmann equation and its variational approximation discussed in Sec.~\ref{theory_scatter}. One can see that for not very large anisotropy ratios, the results of both approaches are comparable, which justifies the use of the variational approach for moderately anisotropic systems. 
  
  Overall, the effective mass anisotropy appears as an important factor affecting the scattering rate.
Along with the observation that the scattering rate is mostly suppressed in anisotropic systems, anisotropy serves as a potential tool for manipulating the transport properties of materials with intrinsic or extrinsically induced anisotropies.

\subsection{\label{sec4d}Electron-phonon scattering}
Another type of scattering processes that are crucially important for semiconductors is the electron-phonon scattering. For simplicity, we limit ourselves to the case of scattering by acoustic phonons and parabolic energy dispersion of electrons. For isotropic 2D materials with parabolic energy dispersion, the most simple way to estimate phonon-limited carrier mobility $\mu=e\tau/m$ for degenerate electron gas ($E_\mathrm{F}\gg k_B T$) is to employ the so-called Takagi formula \cite{Takagi1994}
\begin{equation}
\label{Takagi}
    \mu = \frac{e\hbar^3C}{k_BTm^2g^2},
\end{equation}
where $k_BT$ is the temperature in energy units, $C$ is the elastic constant of a 2D material, $m$ is the effective mass of charge carriers, and $g$ is the deformation potential quantifying the strength of the electron-phonon coupling in the lowest order in atomic displacements. It is tempting to generalize Eq.~(\ref{Takagi}) to the anisotropic case simply by replacing $C$, $m$, and $g$ with their respective direction-dependent counterparts. Such an approach was used in early works on BP to estimate its mobility \cite{Qiao2014}. However, this approach is formally invalid as it ignores momentum dependence of the scattering matrices.

\begin{figure}[t]
\centering
\mbox{
\includegraphics[width=0.7\linewidth]{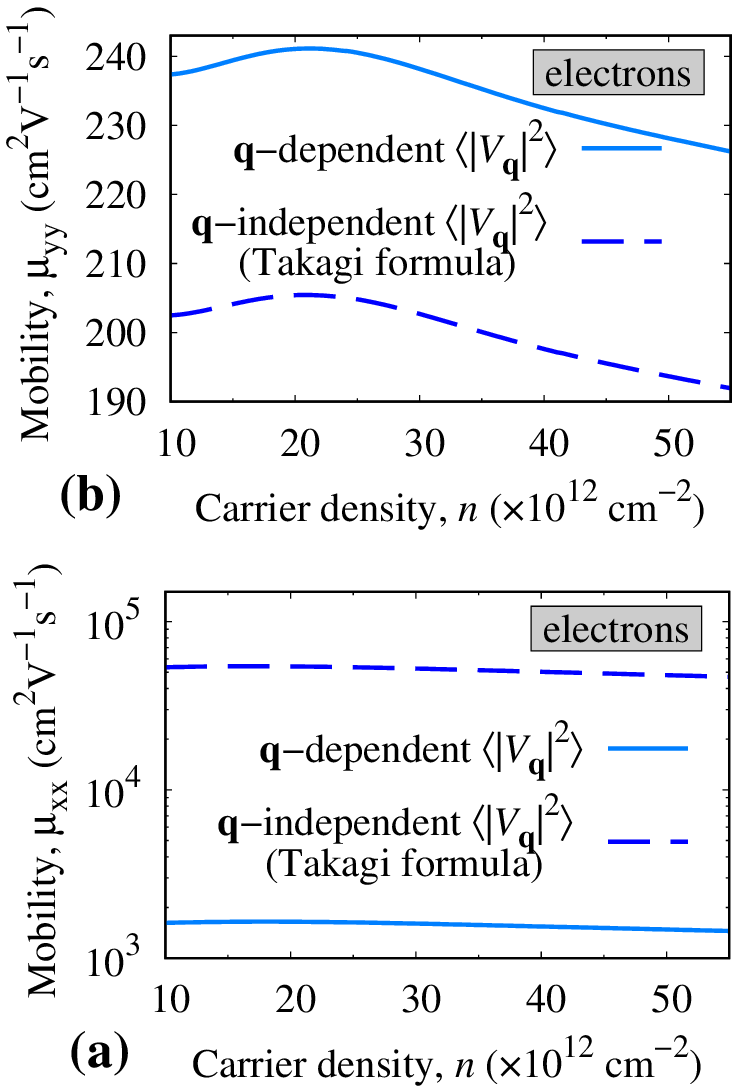}
}
\caption{
Carrier mobility as a function of the electron concentration calculated for monolayer BP along (a) the armchair ($\mu_{xx}$) and (b) zigzag ($\mu_{yy}$) crystallographic directions at the temperature $T=300$ K. The solid and dash curves are obtained using the ${\bf q}$-dependent [Eq.~(\ref{V_phonons})] and ${\bf q}$-independent scattering matrix $\langle |V_{\bf q}|^2\rangle$. The latter case corresponds to the Takagi formula, Eq.~(\ref{Takagi}). The calculations are performed using the parameters from Ref.~\onlinecite{Rudenko2016}, assuming the scattering on in-plane phonons only. Note the logarithmic scale for $\mu_{xx}$ in (a).
}
\label{fig_mobility}
\end{figure}

A more systematic approach to treat the interaction of charge carrier with acoustic phonons in anisotropic 2D systems has been proposed by us in Ref.~\onlinecite{Rudenko2016}. In
the long-wavelength limit, the effective scattering potential
of charge carriers induced by acoustic phonons $V({\bf r})$
can be written in terms of the diagonal components of the deformation potential tensor $g_{\alpha}$ ($\alpha = x,y$) as
\begin{equation}
    V({\bf r}) = g_x u_{xx}({\bf r}) + g_y u_{yy}({\bf r}),
\end{equation}
where
\begin{equation}
    u_{\alpha \beta}({\bf r}) = \frac{1}{2}[\partial_{\alpha} u_{\beta}({\bf r}) + \partial_{\beta} u_{\alpha}({\bf r}) + \partial_{\alpha} h({\bf r}) \partial_{\beta} h({\bf r})]
\end{equation}
with $u_{\alpha}({\bf r})$ and $h({\bf r})$ being in-plane and out-of-plane displacement fields, respectively. In the absence of out-of-plane deformations, the Fourier transform of the scattering potential is given by
\begin{equation}
V_{\bf q} = ig_{x}u_x({\bf q})q_{x} + ig_{y}u_y({\bf q})q_{y},
\end{equation}
where $u_{\alpha}({\bf q})$ are the Fourier components of the displacement field $u_{\alpha}({\bf r})$. Fluctuations of the displacement fields in 2D crystals with orthorhombic symmetry can be described in harmonic approximation by the Hamiltonian
\begin{equation}
    H = \frac{1}{2}\int d^2{\bf r} [C_{11}u^2_{xx} + C_{22}u^2_{yy} + 2C_{12}u_{xx}u_{yy} + 4C_{66}u^2_{xy}], 
\end{equation}
where $C_{\alpha \beta}$ are elastic constants in the Voigt notation. With this Hamiltonian, the thermal average $\langle |V_{\bf q}|^2 \rangle$ can be integrated out exactly. If phonons are considered classically, i.e. when $\hbar \omega({\bf q}\approx {\bf k}_{\mathrm{F}}) \ll k_B T$, the result is \cite{Rudenko2016}
\begin{widetext}
\begin{equation}
\langle |V_{{\bf q}}|^2 \rangle = k_BT \frac{C_{66}({g}_x^2q_x^4+{g}_y^2q_y^4-2{g}_x{g}_yq_x^2q_y^2) + (C_{22}{g}_x^2+C_{11}{g}_y^2-2C_{12}{g}_x{g}_y) q^2_xq^2_y} {C_{66}(C_{11}q_x^4+C_{22}q_y^4-2C_{12}q_x^2q_y^2)+(C_{11}C_{22}-C_{12}^2)q_x^2q_y^2}.
\label{V_phonons}
\end{equation}
\end{widetext}
One can see that the scattering matrix exhibits a sophisticated dependence on both the direction and the magnitude of the scattering wave vector ${\bf q}$. In the isotropic case $C_{11}=C_{22}$ and $C_{12}=C_{11}-2C_{66}$, thus after some algebra Eq.~(\ref{V_phonons}) simplifies remarkably to $\langle |V_{{\bf q}}|^2 \rangle = k_BTg^2/C_{11}$, which is ${\bf q}$ independent. One can see that in this case we immediately recover the Takagi formula by noting that $\tau^{-1} = m\langle |V_{\bf q}|^2 \rangle /\hbar^3$, which directly follows from Eq.~(\ref{tauxx-variational}). Due to the complicated ${\bf q}$ dependence of $\langle |V_{{\bf q}}|^2 \rangle$, the scattering rate cannot be calculated analytically in the anisotropic case.  The importance of the ${\bf q}$ dependence is shown in Fig.~\ref{fig_mobility}, where we compare the electron mobility in monolayer BP calculated at $T=300$~K based on the full ${\bf q}$-dependent scattering matrix Eq.~(\ref{V_phonons}) with the Takagi formula Eq.~(\ref{Takagi}) employing the same set of parameters from Ref.~\cite{Rudenko2016}. Along the zigzag ($y$) direction, both methods yield comparable electron mobility with the difference around $\sim$20\%. Along the armchair direction, the situation is dramatically different: The Takagi formula overestimates the mobility by almost two orders of magnitude. This behavior is due to relatively small deformation potential along the $x$ direction in BP, giving rise to the anisotropy ratio $g_x/g_y \sim 0.1$, meaning $\mu_{xx}/\mu_{yy} \sim (g_y/g_x)^2 \sim 100$ according to the Takagi formula. Such consideration is misleading because the scattering in the $x$ direction does not only depend on the corresponding deformation potential $g_x$, but exhibits a more involved angular dependence described by Eq.~(\ref{V_phonons}). Therefore, the Takagi formula has limited applicability to anisotropic systems, and has to be used with caution.

The approach outlined above is restricted to the scattering by acoustic in-plane phonons with the linear dispersion $\omega^{\mathrm{in}}_{\bf q}\sim q$, i.e., the resulting carrier mobility should only be considered as an upper limit. In 2D systems, charge carrier can also scatter on flexural phonons \cite{Katsnelson-Book} with the quadratic dispersion, $\omega^{\mathrm{flex}}_{\bf q} \sim q^2$. For crystals with mirror (or equivalent) symmetries, single-phonon scattering on flexural phonons is not relevant \cite{Fischetti2016}. In this situation, only two-phonon processes can contribute to the scattering on flexural phonons. These processes were shown to be crucially important for the temperature dependence of electron mobility in free-standing graphene \cite{Katsnelson-Book}. In the majority of cases, however, such processes are considerably weaker than the single-phonon processes considered above and may be neglected in practical calculations of 2D semiconductors \cite{Rudenko2016,Rudenko2019}. 

Apart from acting as a direct scattering source, two-phonon processes involved flexural phonons can lead to a renormalization of the electronic spectrum in doped semiconductors, giving rise to the formation of fluctuation-induced tails at the edges of the bands; the states at the edges can be described as almost localized self-trapped states (flexurons) \cite{Katsnelson2010}. The effect is more pronounced in strongly anisotropic systems, due to quasi-1D enhancement of the electron DOS near the band edges. In the context of doped monolayer BP, this effect was considered quantitatively by Brener \emph{et al.} \cite{Brener2017}.

Optical phonons represent another source of scattering, which becomes important at the temperatures $k_B T \gtrsim \hbar \omega^{\mathrm{opt}}$, where $\omega^{\mathrm{opt}}$ is the characteristic frequency of optical phonons. Due to the large number of optical phonons in real 2D materials, it is difficult to build quantitative analytical theory to capture optical modes. The standard approach to handle this problem is to use \emph{ab initio} calculations based on, e.g., density functional perturbation theory, where electron-phonon matrix elements are evaluated from the first derivative of the effective electronic potential over atomic displacements \cite{Ponce2020}.
Using such methods it has been shown that optical phonons in BP provide an appreciable contribution to carrier mobility at room temperature \cite{Liao2015}. 
In the context of few-layer BP, Gaddemane \emph{et al.} \cite{ Gaddemane2018} have critically examined physical models and computational approaches to electronic transport. They emphasized the importance of the so-called wavefunction-overlap effect, i.e. dependence of the scattering matrix on the initial and final scattering states. They argued that this effect affects not only the magnitude, but also angular dependence of the electron-phonon scattering matrices, i.e., may provide additional contribution to the mobility anisotropy.

\subsection{\label{sec4e}Superconductivity in anisotropic systems}

If the system demonstrates superconducting phase transition, the anisotropy of the electron spectrum in the normal phase should also manifest itself in the properties of the superconducting phase. Particularly, one can expect direction dependence of the superconducting gap along the Fermi surface. From the theory point of view, anisotropic gap is not uncommon for conventional superconductors and has been reported, e.g., for MgB$_2$ \cite{Mazin2001,Choi2002} as well as for heavily doped graphene \cite{Margine2014,Margine2016} as a direct consequence of the Fermi surface anisotropy. Experimentally, however, such anisotropy is not easy to measure due to limited resolution of the most straightforward methods such as ARPES \cite{RevModPhys.93.025006} and smallness of the gap in conventional, that is, not high-temperature, superconductors. Among intrinsically anisotropic crystals, superconductivity has been observed in BP under high pressure as a result of structural transformation \cite{KAWAMURA1985775,XiangLi2018}, as well as in BP doped by alkali atoms at ambient pressure \cite{Zhang2017}. Despite high interest to BP, we are not aware of direct measurements of the superconducting gap in BP.

\begin{figure}[tbp]
\centering
\mbox{
\includegraphics[width=1.0\linewidth]{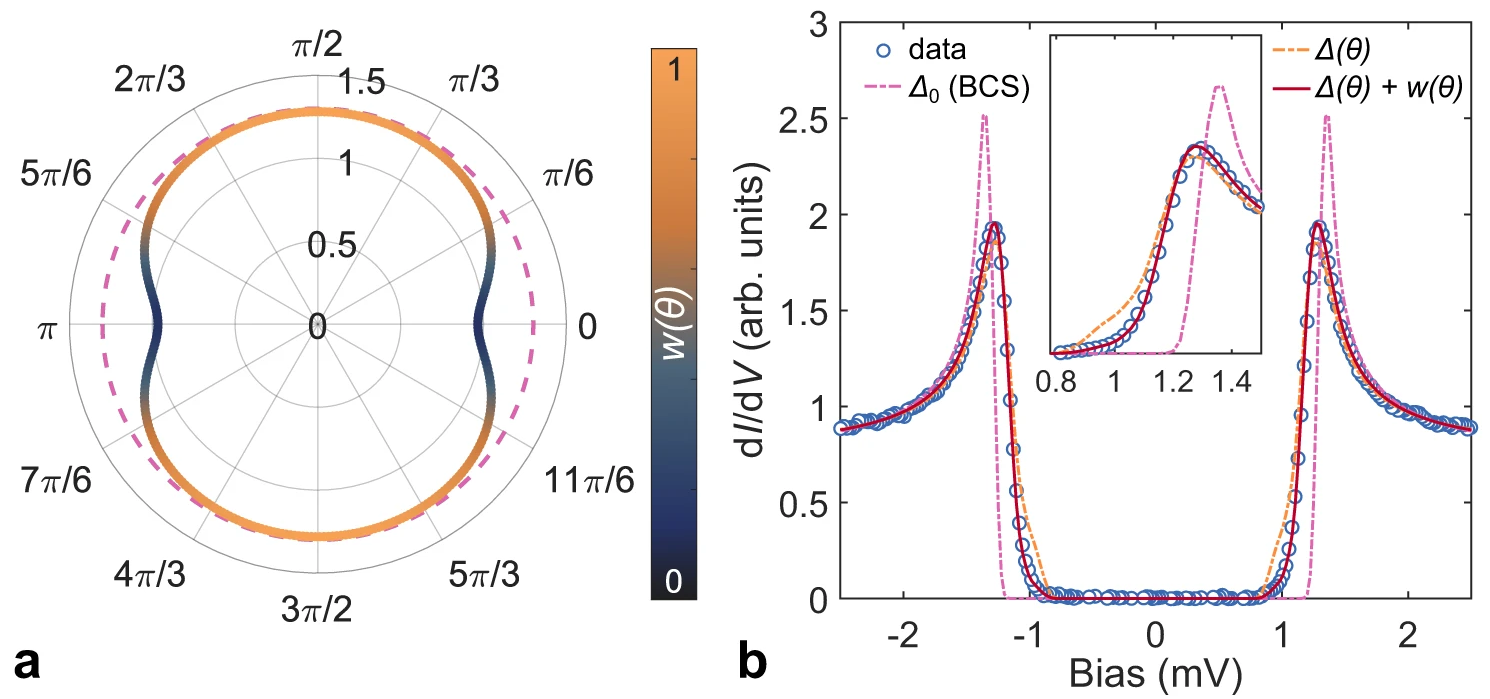}
}
\caption{(a) Polar plot of the anisotropic superconducting gap $\Delta(\theta)$ derived from the Pb/BP model outlined in the text (solid line) shown in comparison with the isotropic gap $\Delta_0$ (dashed line), (b) d$I$/d$V$ STM spectrum measured at $T=30$ mK on a 7-layer Pb film on BP (circles) and results of the model DOS calculations (solid line). The magenta and orange curves represent simulated spectra with an isotropic gap (dashed line in a) and an anisotropic gap without incorporating the weight function $w$, respectively. The latter describes spectral density transfer from superconducting to nonsuperconducting states.
Adapted figure from Ref.~\onlinecite{Kamlapure2022}.
}
\label{fig6_superconductivity}
\end{figure}

Another approach to study anisotropy of the superconducting properties is to induce anisotropy in otherwise isotropic superconductors utilizing the proximity effects.
In this context, it is worthwhile to mention a recent paper of Kamlapure \emph{et al.} \cite{Kamlapure2022} where superconductivity of a few-layer Pb film on the surface of BP was studied using STM techniques. Here, STM experiments demonstrate that BP substrate significantly renormalizes the superconductivity in Pb film, and clearly reveal anisotropic structure of superconducting vortex cores. Also, the d$I$/d$V$ STM spectra in the superconducting phase indicate angular dependence of the superconducting gap (see Fig.~\ref{fig6_superconductivity}). 

There are multiple mechanisms that could be relevant for the induced anisotropy of superconductivity. Among them are anisotropy of the dielectric screening which can result in the anisotropy of the Coulomb pseudopotential \cite{Allen1983,vonsovsky_superconductivity}, as well as angular dependence of the electron-phonon interaction due to anisotropic phonons in the substrate. 
However, it turns out that the main source of anisotropy in the Pb/BP system is the hybridization of electron states of BP and Pb at the interface. 
The authors of Ref.~\onlinecite{Kamlapure2022} have employed the following minimal two-band model to simulate this effect. The model is described by the Hamiltonian:
\begin{eqnarray}
\notag
    H = \sum_{{\bf k}\sigma}\xi_{\bf k}c^{\dag}_{{\bf k}\sigma}c_{{\bf k}\sigma} + \sum_{{\bf k}{\bf k}'}V_{{\bf k}{\bf k}'}c^{\dag}_{{\bf k}\uparrow}c^{\dag}_{-{\bf k}\downarrow}c_{{\bf k}'\downarrow}c_{{\bf k}'\uparrow} \\ + t\sum_{{\bf k}\sigma}\left[ c^{\dag}_{{\bf k}\sigma}b_{{\bf k}\sigma} +\text{h.c.}\right] + \sum_{{\bf k}\sigma} \eta_{{\bf k}} b^{\dag}_{{\bf k}\sigma}b_{{\bf k}\sigma}.
    \label{sc-model}
\end{eqnarray}
Here, $c^{\dag}_{{\bf k}\sigma}$ ($c_{{\bf k}\sigma}$) and $b^{\dag}_{{\bf k}\sigma}$ ($b_{{\bf k}\sigma}$) are creation (annihilation) operators of electrons in Pb and BP, respectively, $\xi_{{\bf k}}$ and $\eta_{{\bf k}}$ represent the isotropic and anisotropic dispersion of electrons in Pb and BP, respectively, and $t$ describes a hybridization between the Pb and BP states. $V_{{\bf k}{\bf k}'}$ is a function, in spirit of BCS model, describing the Cooper pairing of the Pb states. After mean-field decoupling of the second term in Eq.~(\ref{sc-model}), one can employ the Nambu-Gorkov formalism and define the matrix Green's function $G = (i\omega_n - H)^{-1}$ as an inverse of the matrix
\begin{equation}
G^{-1} = 
\begin{pmatrix}
    i\omega_n - \xi_{\bf k} & -\Delta_{\bf k} & -t & 0 \\
    -\Delta^*_{\bf k} & i\omega_n + \xi_{\bf k} & 0 & t \\   
    -t & 0 & i\omega_n - \eta_{\bf k}& 0 \\    
    0 & t & 0 & i\omega_n + \eta_{\bf k}   
\end{pmatrix},
\label{G-matrix}
\end{equation}
where $\omega_n$ is the fermionic Matsubara frequency, and $\Delta_{\bf k}=\sum_{{\bf k}'}V_{{\bf k}{\bf k}'}\langle c_{-{\bf k}'\downarrow} c_{{\bf k}'\uparrow}\rangle$ plays the role of the superconducting gap function. The average is calculated by inverting Eq.~(\ref{G-matrix}) and taking the anomalous part of the Green's function matrix, namely, $\langle c_{-{\bf k}'\downarrow} c_{{\bf k}'\uparrow}\rangle = T\sum_{i\omega_n}G_{12}(i\omega_n)$, which defines an implicit equation for ${\Delta}_{\bf k}$. In the superconducting phase, the quasiparticle dispersion is then given by
\begin{eqnarray}
\notag
    \lambda^2_{\pm}({\bf k}) = \frac{1}{2}\left[ 2t^2 + E^2_{{\bf k}} + \eta^2_{\bf k} \pm \right. \quad \quad \quad \quad \quad \quad \quad \quad \\
    \left. \sqrt{ (E^2_{\bf k} - \eta^2_{\bf k} )^2 +4t^2[\Delta^2_{\bf k} + (\eta_{\bf k} + \xi_{\bf k})^2] } \right],
    \label{lambda-pm}
\end{eqnarray}
where $E_{\bf k}=\sqrt{\xi^2_{\bf k} + \Delta^2_{\bf k}}$. Importantly, the hybridization $t$ affects both the superconducting gap and the quasiparticle dispersion. Although the pairing potential is applied exclusively to the Pb states, the superconducting state is also determined by the anisotropic BP states, yielding anisotropic quasiparticle dispersion even for constant (isotropic) gap. Another important effect is the spectral density transfer from superconducting to non-superconducting states 
(in Fig. \ref{fig6_superconductivity}, this effect is shown by incorporating the weight function $w$ defined in Ref.~\onlinecite{Kamlapure2022}). 

In the weak hybridization limit ($t \ll \eta({\bf k}_\mathrm{F})$), 
the gap can be expressed in a closed-form as
\begin{equation}
    \Delta^2_{\bf k} \simeq \Delta^2_0\left[ 1- \frac{\eta_{\bf k}t^2}{[\eta_{\bf k} - \xi_{\bf k}]^2[\eta_{\bf k} + \xi_{\bf k}]} \right].
\end{equation}
The authors of Ref.~\onlinecite{Kamlapure2022} have considered specific energy dispersion of the form $\xi_{\bf k} = \hbar^2k^2/2m^* - \mu$ and $\eta_{\bf k} = \hbar^2k^2F(\theta)/2m_x + \delta - \mu$, where $F(\theta)=1+\epsilon\,\mathrm{sin}^2(\theta)$ with $\epsilon=m_x/m_y-1$, $\mu$ is the chemical potential, and $\delta$ is an offset between the Pb and BP bands. With this dispersion, the angular dependence of the gap at the Fermi surface takes the form,
\begin{equation}
\Delta(\theta) = \Delta_0 \left[ 1 - \frac{t^2}{[\delta - \mu + \frac{m^*}{m_x}F(\theta)\mu]^2} \right]
\label{delta-theta}
\end{equation}
which is depicted in Fig.~\ref{fig6_superconductivity}(a), clearly demonstrating anisotropic behavior. Fig.~\ref{fig6_superconductivity}(b) shows that experimental data can be nicely fitted to the model. This suggests that the degree of electronic anisotropy of the underlying substrate can be used to tune superconducting properties of 2D and quasi-2D materials. It is interesting to note that the gap renormalization can take place even for small hybridization $t$. From Eq.~(\ref{delta-theta}) one can see that for strongly anisotropic systems in the regime $\delta \sim \mu$ the behavior of $\Delta(\theta)$ can change significantly, and $\Delta(\theta)$ might even approach zero at certain $\theta$. Although such regime is unlikely to be realized in typical superconducting metals, it might exist in highly doped semiconductors. This seems to be an interesting direction of future research.

\section{Summary and outlook\label{outlook}}
In this manuscript, we presented an overview of anisotropy effects in 2D materials, focusing on the electronic, optical, and transport properties. Starting from basic models to the electronic structure, we showed how anisotropy comes into play in the formation of properties of semiconductors, Dirac, and semi-Dirac materials. Apart from purely quantitative effects affecting the numerical coefficients, anisotropy turns out to be responsible for a number of qualitative effects that have no analogs in isotropic systems. Such effects include, for example, unusual propagation of electromagnetic waves (hyperbolic plasmons) as well as modification of the scaling laws.

Among 2D anisotropic materials, semiconductors are the most studied class of materials thanks to the diversity of experimentally available materials of this kind. Anisotropic Dirac and semi-Dirac materials are less ubiquitous, but display diversity of interesting characteristics that call for experimental verification. At the same time, such materials demonstrate enhanced many-body effects (e.g., Fermi velocity renormalization) which appear even at the lowest level of theory. Some direction-dependent physics of this kind (e.g., anisotropic superconductivity) can be induced even in 2D isotropic systems by utilizing the proximity effects. This opens new avenues for future research, both theoretical and experimental.

In our review, we did not discuss anisotropic aspects of 2D magnetism, which is a promising, rapidly growing field of research. In-plane anisotropy of 2D magnets offers a complicated phase diagram even at the level of simple spin models. The presence of spin-orbit coupling in conjunction with a quasi-1D character of the electronic subsystem may give rise to a number of interesting phenomena in the domain of spintronics. We believe further research in this direction will lead to unexpected and exciting results.

\begin{acknowledgements}
The work was supported by European Research Council via Synergy Grant 854843 - FASTCORR.
\end{acknowledgements}

\bibliography{ref}

\end{document}